\documentclass[fleqn,usenatbib]{mnras}

\usepackage{newtxtext,newtxmath}

\usepackage[T1]{fontenc}

\DeclareRobustCommand{\VAN}[3]{#2}
\let\VANthebibliography\thebibliography
\def\thebibliography{\DeclareRobustCommand{\VAN}[3]{##3}\VANthebibliography}

\usepackage{graphicx}	
\usepackage{amsmath}	

\graphicspath{{./}{figures/}}

\title[IGL in a GAMA group at $z\sim0.21$]{Galaxy and Mass Assembly (GAMA): Extended Intra-Group Light in a group at $z=0.2$ from deep Hyper-Suprime Cam images}

\author[Mart{\'i}nez-Lombilla et al.]{Cristina Mart{\'i}nez-Lombilla,$^{1,2}$\thanks{E-mail: c.martinezlombilla@unsw.edu.au}
Sarah Brough,$^{1}$
Mireia Montes,$^{3}$\thanks{STScI Prize fellow}
Roberto Baena-Gall{\'e},$^{4}$
\newauthor Mohammad Akhlaghi,$^{5}$
Ra{\'u}l Infante-Sainz,$^{5}$
Simon P. Driver,$^{6}$
Benne W. Holwerda,$^{7}$
\newauthor Kevin A. Pimbblet,$^{8}$
and Aaron S.G. Robotham$^{6}$
\\
$^{1}$ School of Physics, University of New South Wales, Sydney, NSW 2052, Australia\\
$^{2}$ Australian Research Council, Centre of Excellence for All-Sky Astrophysics in 3 Dimensions (ASTRO 3D), Australia\\
$^{3}$ Space Telescope Science Institute, 3700 San Martin Drive, Baltimore, MD 21218, USA\\
$^{4}$ Universidad Internacional de la Rioja. Avenida de la Paz, 137, 26006 Logroño, La Rioja, Spain\\
$^{5}$ Centro de Estudios de Física del Cosmos de Aragón (CEFCA), Plaza San Juan 1, 44001 Teruel, Spain\\
$^{6}$ International Centre for Radio Astronomy Research, University of Western Australia, Crawley, WA 6009, Australia\\
$^{7}$ Department of Physics and Astronomy, University of Louisville, Louisville, KY 40292, USA\\
$^{8}$ E.A.Milne Centre for Astrophysics, University of Hull, Cottingham Road, Kingston-upon-Hull, HU6 7RX, UK
}

\date{Accepted 2022 October 21. Received 2022 October 21; in original form 2022 September 1}

\pubyear{2022}

\begin{document}
\label{firstpage}
\pagerange{\pageref{firstpage}--\pageref{lastpage}}
\maketitle

\begin{abstract}
We present a pilot study to assess the potential of Hyper Suprime-Cam Public Data Release 2 (HSC-PDR2) images for the analysis of extended faint structures within groups of galaxies. We examine the intra-group light (IGL) of the group 400138 ($M_{\rm{dyn}}= 1.3 \pm 0.5 \times 10^{13} $M$_{\sun}$, $z\sim 0.2$) from the Galaxy And Mass Assembly (GAMA) survey using Hyper-Suprime Cam Subaru Strategic Program Public Data Release 2 (HSC-PDR2) images in $g$, $r$, and $i$ bands. We present the most extended IGL measurement to date, reaching down to $\mu _{g}^{\rm{lim}}=30.76$~mag~arcsec$^{-2}$ ($3 \sigma$; $10 \times 10$~arcsec$^{2}$) at a semi-major axis of 275~kpc. The IGL shows mean colour values of $g-i=0.92$, $g-r=0.60$, and $r-i=0.32$ ($\pm$0.01). The IGL stellar populations are younger ($2-2.5$~Gyr) and less metal-rich ([Fe/H]~$ \sim -$0.4) than those of the host group galaxies. We find a range of IGL fractions as a function of total group luminosity of $\sim 2-36 \%$ depending on the definition of IGL, with larger fractions the bluer the observation wavelength. The early-type to late-type galaxy ratio suggests that 400138 is a more evolved group, dominated by ETGs, and the IGL fraction agrees with that of other similarly evolved groups. These results are consistent with tidal stripping of the outer parts of Milky Way-like galaxies as the main driver of the IGL build-up. This is supported by the detection of substructure in the IGL towards the galaxy member 1660615 suggesting a recent interaction ($<1$~Gyr ago) of that galaxy with the core of the group.

\end{abstract}





\section{Introduction} \label{sec:intro}

In dense environments such as groups or clusters of galaxies, a significant number of stars initially belonging to each galaxy are ripped out with every interaction within that group. These stars congregate in an extended faint light component, not bound to any individual galaxy but to the system potential, forming the intra-group and intra-cluster light (IGL and ICL; see \citealt{Mihos2019, Contini2021} and \citealt{Montes2022} for reviews). As a result, these faint components form a fossil record of all the dynamic processes a group or cluster has undergone  \citep[e.g.][]{Merritt1984}. Unveiling the quantity of IGL and its stellar population properties provides a holistic view of the system's interaction and mass assembly history. However, the diffuse and faint nature of both the IGL and ICL ($\mu _{V}^{\rm{IGL}}>26.5$~mag~arcsec$^{-2}$; \citealt[][]{Rudick2011}) makes their study a technical challenge and its physical properties are not well established yet. 

Galaxy groups are the perfect systems for the study of processes involved in the mass assembly history of dense systems of galaxies. According to the $\Lambda$CDM scenario, galaxies spend most of their lives in groups, and these groups of galaxies form the building blocks of more massive and complex galaxy clusters \citep{DeLucia2006}. Therefore, the IGL is a crucial piece in understanding the growth of the ICL \citep[e.g.][]{Willman2004, Fujita2004, Gnedin2003, SommerLarsen2006, Rudick2006, Rudick2009, Bahe2013, Contini2014, Bahe2017, Remus2017, Pillepich2018, Canas2020}. Using a semi-analytic model of galaxy formation, \citet{Contini2014} showed that up to $30 \%$ of the ICL in simulations comes from the ``pre-processing'' of this light in groups and clusters of galaxies. Observational works have also found that the diffuse light component associated with groups of galaxies could be an important component of the ICL in clusters \citep{Krick2006, Krick2007, Mihos2017}. This is good evidence for the importance of studying the diffuse light in groups of galaxies.




Because the IGL is very faint we are only just starting to understand its nature. Only a few works have detected and measured the IGL component and these are all of individual or small samples of groups of galaxies. These IGL detections are, in general, from dedicated observations of dense, strongly interacting groups like compact groups \citep[e.g.][]{Aguerri2006, DaRocha2005, DaRocha2008, Durbala2008, Poliakov2021, Ragusa2021}, with only a few studies of loose groups \citep{CastroRodriguez2003, Spavone2018, Cattapan2019, Iodice2020, Raj2020}. This is because the IGL is easier to observe in compact groups than in loose groups because galactic interactions and intense stripping are very common in these systems and these are highly efficient mechanisms in the build-up of the diffuse light component. Even more limited is the knowledge of the IGL properties at intermediate redshifts ($0.1 \lesssim z \lesssim 0.6$). Diffuse light detections at these redshifts have only been made for massive groups and clusters \citep[$M > 3 \times 10^{13} $M$_{\sun}$;][]{Gonzalez2007, Gonzalez2013, DeMaio2016, DeMaio2018, DeMaio2020}. 




Measuring the colours of the galaxies and the faint component is key to discern which mechanisms play the largest roles in the build-up of the IGL. In fact, depending on the IGL radial colour gradients, we can infer its potential formation scenario: the complete disruption of dwarf galaxies or the partial tidal stripping of satellite galaxies produce a radial colour gradient in the IGL stellar populations, while major mergers produce relatively uniform IGL colour profiles \citep[see][for a more extensive explanation]{DeMaio2018}. 

The IGL, as a low surface brightness (LSB) feature, presents technical challenges beyond the pure detection of its faint light. This faint component can be easily contaminated by foreground and background sources, the intrinsic emission of the sky, the presence of Galactic cirrus contamination, reflections along the optical path of the light, electronic issues in the detectors or the atmospheric conditions. To minimise the contribution of any of these undesirable light sources to our target, a complex highly-human-dependent data process is required. However, that process has become a well established set of techniques within the LSB community over the last few decades \citep[e.g.][]{Tyson1998, deJong2008, Slater2009, Duc2015, Capaccioli2015, TrujilloFliri2016, MartinezLombilla2019b, Borlaff2019, InfanteSainz2020, Montes2021a, Trujillo2021}. This process includes the modelling of a very extended point spread function \citep[PSF;][]{Slater2009, Sandin2014, Sandin2015, InfanteSainz2020} to allow a reliable colour analysis of the target as the light scatters differently depending on the observed wavelength. This is of particular relevance for this work.



\begin{figure*}
\centering
\includegraphics[width=12.4cm]{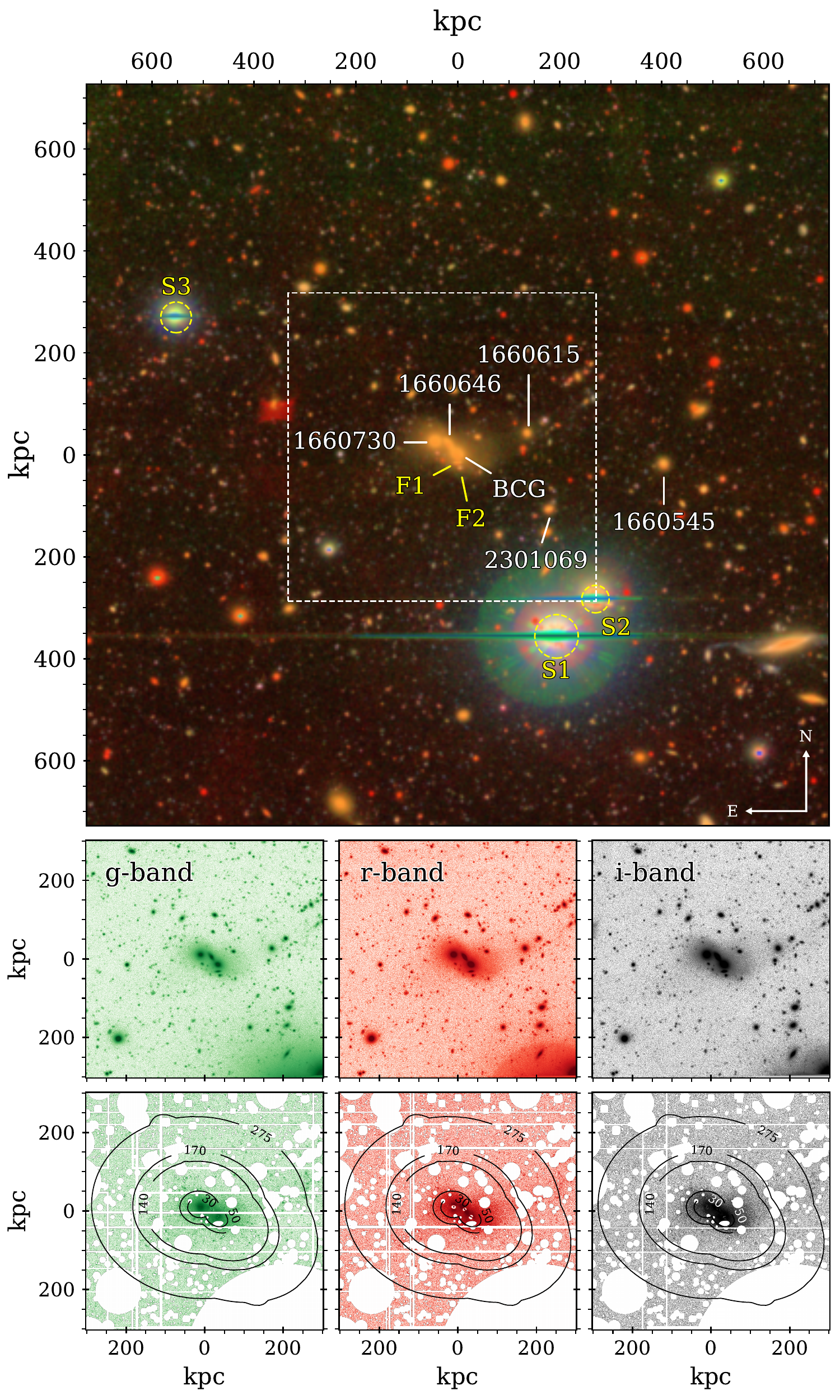}
\caption{\textit{Top}: HSC $g$-$r$-$i$ colour composite image of the 400138 group of galaxies within our $7 \times 7$~arcmin field of view \citep[processed with the algorithm \texttt{make\_lupton\_rgb} described in][]{Lupton2004}. Galaxy group members are labelled with their GAMA ID \citep{Robotham2011}, except for the BCG. The 3 bright stars (S1, S2, and S3) and the foreground galaxies (F1 and F2) that were modelled and removed to avoid their contribution to the group flux are also labelled. \textit{Middle row}: Zoom-in over the group as indicated by the white square in the top panel in each of the $g$ (green), $r$ (red) and $i$ (grey) bands from left to right. \textit{Bottom row}: Zoom-in over the IGL component in each band as above. These images are masked including CCD gaps and bad pixels (white regions), sky subtracted, and the bright stars and core galaxies have been modelled and subtracted to account for the influence of the extended wings of their PSF. The black contours indicate 5 of the profile annuli with radii (in kpc) given in the apertures. These images are sharply colour contrasted to emphasise the faint IGL over the background and the residuals of the galaxy models in the core region.  
  \label{fig:GroupIm}}
\end{figure*}

Our goal is to perform a pilot study to test the potential of Hyper Suprime-Cam Subaru Strategic Program Public Data Release 2 \citep[HSC-SSP PDR2;][]{Aihara2019} images to explore the faint IGL component within groups of galaxies. This work will set the basis for the future analysis of the IGL in a sample of groups of galaxies using HSC data. Our long term goal is to apply the procedures shown in this paper to the upcoming data from the 10-year Legacy Survey of Space and Time (LSST) at the Vera C. Rubin Observatory \citep[][]{Ivezic2019, Montes2019a, Brough2020}. In this study, we aim to contribute to a broader understanding of the formation and evolution properties of the IGL. We have used the Galaxy And Mass Assembly survey \citep[GAMA;][]{Driver2011} Galaxy Group Catalogue \citep[G$^3$Cv10;][]{Robotham2011} and observations from the HSC-PDR2 Subaru/XMM-Newton Deep Survey field \citep[SXDS;][]{Furusawa2008}, the only Ultradeep region where GAMA and HSC-PDR2 overlap, to detect and analyse the IGL in a group of galaxies down to surface brightness limits at 3$\sigma$ within 10~$\times$~10~arcsec$^{2}$ of $\mu _{g} ^{\rm{lim}}=30.76$~mag~arcsec$^{-2}$, $\mu _{r} ^{\rm{lim}}=29.82$~mag~arcsec$^{-2}$, and $\mu _{i} ^{\rm{lim}}=29.41$~mag~arcsec$^{-2}$. In particular, we evaluate the IGL surface brightness radial distribution and colour properties in comparison to those of the group galaxies. We analyse the IGL stellar populations to provide information about the origin of the IGL. To undertake this study, we have developed a semi-automatic IGL detection and colour analysis method, building on previous works on the study of low surface brightness structures \citep[e.g.][]{Montes2014, MartinezLombilla2019a, Montes2021a}. This method is entirely Python-based with multiple iterative steps that build on their previous best result to minimise human interaction. We will apply this method to a significant sample of groups in our follow-up work. In this pilot analysis we focus on the GAMA group identified by the ID 4001389 (RA 35.834163~deg, DEC -5.454157; J2000), also known as XMMXCS~J022318.3-052707.6 or X-CLASS~0346. According to the G$^{3}$Cv10 catalogue \citep[][]{Robotham2011}, 400138 is a loose group of mass $M_{\rm{dyn}}= 1.34 \pm 0.5 \times 10^{13} $M$_{\sun}$ ($v_{\rm{disp}} = 222 \pm 43$~km~s$^{-1}$), located at redshift $z\sim0.21$, and composed of 6 galaxy members. Three of these members are concentrated in the core region, while the other three galaxies are at larger distances (see Figure~\ref{fig:GroupIm}). Although there is a dense region in the core, this group configuration does not satisfy the Hickson criteria \citep{Hickson1997} to be considered as a compact group. 400138 is an intermediate redshift group of galaxies with a visible extended halo around the central galaxies and an interesting dense core environment that favours a potential IGL detection to test our method. In the field of view of 400138 there are three bright stars in an optimal spatial configuration that will challenge our PSF modelling and subtraction, a key step for a reliable colour analysis of any low surface brightness structure. The combination of a potentially clear IGL detection and a challenging PSF correction made galaxy group 400138 the favourite candidate amongst the possible targets for our pilot study.



For the first time, we have detected IGL in a GAMA group at redshift 0.21 (i.e. $\sim 2.5$~Gyr ago) from HSC-PDR2 data. The results of the IGL analysis of the 400138 group of galaxies are presented in this paper as follows: In Section~\ref{sec:data} we present our data and explain our selection criteria; we then explain how we prepared that data in Section~\ref{sec:prepdata}, giving details of key processes such as the sky subtraction, PSF modelling and mask construction. The main results are given in Section~\ref{sec:results}. We discuss these results in Section~\ref{sec:discussion} putting particular focus on the study of the IGL radial distribution, the properties of its stellar population, and the fraction of IGL with respect to the total light of the group in order to disentangle the formation and mass assembly history of the IGL in this group. Finally, we draw our conclusions in Section~\ref{sec:conclusions}.

Throughout this work we adopt a standard cosmological model with the following parameters: $H_{0}=70$~km~s$^{-1}$~Mpc$^{-1}$, $\Omega_{m}=~0.3$, and $\Omega_{\Lambda}=0.7$. All magnitudes are in the AB magnitude system.\\


\section{Data and target selection criteria} \label{sec:data}

\subsection{The Galaxy And Mass Assembly survey (GAMA)} \label{subsec:GAMA}

The galaxy group selection was based on the groups available in the Galaxy And Mass Assembly survey \citep[GAMA,][]{Driver2011} Galaxy Group Catalogue \citep[G$^3$Cv10;][]{Robotham2011}. GAMA is a spectroscopic survey carried out over 210 nights from 2008 to 2014, using the optical AAOmega multi-object spectrograph, installed at the 3.9~m Anglo-Australian Telescope (AAT) in the Siding Spring Observatory (Australia). The survey consists of $\sim$300,000 galaxies down to a depth of $r <19.8$~mag, split into five regions. Our target falls in the GAMA G02 region, the only one where the Ultradeep HSC-SSP PDR2 layer overlaps the GAMA sample \citep{Baldry2018}. GAMA groups are found using a friends-of-friends (FoF) based grouping algorithm \citep{Robotham2011}. The halo mass is estimated assuming that its dynamical mass is proportional to $\sigma^{2}R$, where $\sigma$ is the velocity dispersion of the group, and $R$ is the group radius \citep[Eq.~18 in][]{Robotham2011}. The stellar masses of GAMA group galaxies are estimated from optical photometry using an empirically calibrated colour mass relation based on the relation of the colour $(g - i)$ and the mass-to-light ratio $M_{*}/L_{i}$ \citep{Taylor2011}. The photometic values for each galaxy were measured from Canada-France-Hawaii Telescope Legacy Survey \citep[CFHTLS;][]{Ilbert2006}, which is preferred as it is the deepest available.

\subsection{The Hyper Suprime-Cam Subaru Strategic Program Public Data Release 2} \label{subsec:HSC}

Our images were collected by the Hyper Suprime-Cam \citep[HSC;][]{Miyazaki2018}, at the prime focus of the 8.2~m Subaru Telescope. This facility is operated by the National Astronomical Observatory of Japan on the summit of Maunakea (Hawaii, USA). The HSC is a 1.77~deg$^{2}$ imaging camera with a pixel scale of 0.168~arcsec. It is composed of 116 charge-coupled devices (CCDs; 104 for science, 4 for the auto guider and 8 for focus monitoring). We use data from the HSC Subaru Strategic Program \citep[HSC-SSP;][]{Aihara2018a}, a three-layered survey (Wide, Deep, and UltraDeep) of 1400~deg$^{2}$ in five different broad bands ($grizy$) and four narrow filters. The survey was ongoing when we started this analysis. In particular, our galaxy group is located in the Subaru/XMM-Newton Deep Survey field \citep[SXDS;][]{Furusawa2008}, in the UltraDeep layer of the HSC Public Data Release 2 \citep[hereafter HSC-PDR2;][]{Aihara2019}. We used 3 co-added science images in the three HSC-$G$ ($g$), HSC-$R$ ($r$), and HSC-$I$ ($i$) bands, with surface brightness limits of 28.5, 28.1, and 27.9~mag~arcsec$^{-2}$, respectively \citep{Aihara2019}.   

Throughout the development of this project the HSC Public Data Release 3 \citep[HSC-PDR3;][]{Aihara2022} became available. This is an updated and complete version in terms of observation time of the HSC-PDR2. However, we did not use this data for this pilot due to the differences in the sky conditions, the observing strategy, and data treatment pipelines introduce changes that should be accounted in the extended PSFs. This implies the construction of a new extended PSF in each band, equivalent to months of work. On the other hand, the HSC-PDR2 fulfils the requirements for our pilot study and has been widely tested for low surface brightness studies \citep[e.g.][]{Huang2018, Huang2020, Li2022}. In a follow up work where we will analyse the IGL in a large sample of galaxy groups we will evaluate the potential of the HSC-PDR3 data in the faint regime around extended objects.

\subsection{Target selection criteria} \label{subsec:selcrit}

Our selection criteria was based on the goal of performing a pilot study of the IGL properties in GAMA groups using HSC images. We aim to prove that we can detect the faint IGL in groups of galaxies in HSC-SSP data in order to allow for further study of a larger sample of systems. First, we required our target to be located in the UltraDeep SXDS region, the only Ultradeep HSC-PDR2 layer that overlaps with GAMA data. In order to guarantee the reliability of the group properties from the G$^3$Cv10 catalogue, we only preselect systems with more than 4 galaxy members, whose uncertainties in the velocity dispersion measurements are below $40\%$, and with a group radius (defined by the 50th percentile group member) to be lower than 0.7~Mpc/h so we avoid very extended groups. The selected candidate should also be a system with spectroscopic redshift $0.19 < z < 0.50$. This redshift range ensures extended galaxy profiles can be reliably extracted up to $\sim 300$~kpc with no over- or under-subtraction due to residual background issues in the HSC-PDR2 data \citep[][]{Huang2018, Li2022}. Also, these systems are good candidates to understand the still poorly explored IGL properties at intermediate redshift. This selection provided a sample of 43 groups of galaxies. For the particular goals of this pilot study, we looked for groups without obvious recent interactions or bright tidal tails or shells. In addition to the above criteria, the group 400138 shows a clear visible extended halo around the brightest cluster galaxy (BCG) and core region that resembled an IGL component which enables us to evaluate our detection methods. By choosing a system with a likely large fraction of IGL distribution we will not fully demonstrate the potential of our method to detect IGL in a group with a low fraction of IGL. However, in this first work, we focus on the reliability of the complex procedures such as the PSF modelling and subtraction, or the analysis of the IGL stellar populations.

In terms of the field of view, the 400138 group is surrounded by three very bright foreground stars, two of them (the brightest: S1 and S2 in Figure~\ref{fig:GroupIm}) located at close angular separation from the core of the group and also very close to each other. This last point is a challenge in terms of the complex process of PSF modelling and subtraction of both bright stars and galaxies (see Section~\ref{subsec:scatlight}). Thus, the proximity between S1 and S2 allow us to test an extreme situation where the fluxes of both stars overlap and, at the same time, are contaminating the flux of the core region of the group. By selecting the 400138 group as our pilot target we are testing the quality of our PSF modelling and removal techniques, both critical for a reliable IGL colour analysis. In summary, the physical properties of the galaxy group 400138 allow for a potentially clear IGL detection avoiding known background subtraction issues, while its spatial configuration within the field of view ensures a test of our PSF correction for a reliable stellar population study. These reasons led to the group 400138 being the preferred candidate among the possible targets for our pilot study. 

        

\section{Preparation of the data} \label{sec:prepdata}

The top panel in Figure~\ref{fig:GroupIm} shows an HSC colour-composite image of the cutout region of 7~arcmin$^{2}$ around our group of galaxies. There we see a diffuse stellar envelope around the three galaxies located at the centre of the system, this is the BCG, 1660730, and 1660646. This faint light also sprawls towards 1660615 in the north direction showing a tail-like shape. According to a visual check on the data, on the unsharped images (this is, the original group images minus the same image convolved with a Gaussian filter), and on the 2D models (see Sections~\ref{subsec:mask} and \ref{subsubsec:2Dmodels} respectively), all the galaxies in the group show a spheroidal structure with the exception of 1660646 that has a disk.

The goal of this work is to analyse the IGL of the 400138 group of galaxies. We aim to estimate both the amount of IGL in the system, and the nature of its stellar populations. Prior to that, we need to prepare the data for the study of these low surface brightness structures. These procedures are optimised to avoid contamination of our sources of interest from either galactic cirrus, foreground and background sources, or due to atmospheric and telescope/instrument effects. In terms of software, these preparatory steps are fully Python based, the only exception being the 2D model fitting of the galaxies (details in Section~\ref{subsubsec:2Dmodels}) for which we used \textsc{imfit} \citep{Erwin2015}. The detailed explanations of the tools, methods and parameters used are in the following sections.

\subsection{Galactic cirrus emission} \label{subsec:cirrus}

The dust from the local interstellar medium diffuses light from Milky Way stars, affecting astronomical images. This is called Galactic cirrus emission. When analysing deeper images, Galactic cirrus can contaminate the data as it resembles the shape and brightness of low surface brightness structures \citep[e.g.][]{Chiboucas2009, Cortese2010, Chiboucas2013, Duc2018, Barrena2018, RamirezMoreta2018, Roman2020}. 

The HSC-SSP fields are chosen to be low in Galactic cirrus and we also checked this for our region of interest. As the peak of this dust emission falls in the far infrared due to its low temperature \citep{Low1984, Veneziani2010}, we examined 100~micron images from the Infrared Astronomical Satellite \citep[IRAS;][]{Neugebauer1984} Sky Survey Atlas\footnote{Data downloaded from the NASA/IPAC Infrared Science Archive \url{https://irsa.ipac.caltech.edu/applications/IRAS/ISSA/}}. Far infrared maps trace the regions where extended cirrus structures contribute to the optical images. However, Galactic cirrus can also be compact and small, so undetectable with the IRAS angular resolution and could be confused with low surface brightness galaxies. This is a potential problem to be aware of.

The IRAS images of our field of view suggested that the group is located in a region of the sky with low dust contamination. We find a maximum flux of 2.8~MJy/sr, which allows for a reliable colour analysis of our system. Furthermore, the Galactic extinction correction \citep{Schlafly2011} also returns low values: $A(g)$=0.092, $A(r)$=0.066, $A(i)$=0.050. We use these values throughout this work.

\subsection{Sky subtraction and extended radial profiles extraction} \label{subsec:skySBP}

The background estimation is the most important systematic effect when studying the outskirts of galaxies. As a consequence of this, the sky subtraction remains a potentially large systematic uncertainty. In this section we detail our approach to minimise this important source of error.

The data from the HSC-SSP PDR2 is already co-added and sky-subtracted. The global sky subtraction algorithm used in PDR2 is an improved version of that used in the PDR1 data release \citep[][]{Aihara2018}. It now preserves the wings of extended objects allowing for more reliable studies of extended halos (see details in \citealt[][]{Aihara2019}; e.g. \citealt{Huang2018, Li2022}). This is crucial for performing a study of low surface brightness sources of any kind. To check this sky subtraction we studied the sky around our group of galaxies to re-evaluate and correct --if required-- for any remaining sky emission or over-subtraction due to either the original HSC-SSP PDR2 sky subtraction or our later PSF treatment.  

We based our second-order sky subtraction method on the ideas and steps in \cite{PohlenTrujillo2006} and in \citet[][]{MartinezLombilla2019a}: measuring the sky level from the outer and flat points of very extended radial profiles of the BCG. However, our group does not have a clear massive brightest group member, but three core galaxies of similar sizes and luminosities where the brightest one has been identified as the BCG \citep[][]{Robotham2011}. Thus, this BCG is not the only contributor to the central faint light of the system. As a result, we treat these core members (the BCG, 1660730, and 1660646) as a whole system that accounts for the centre of mass (or luminosity) of the group. The sky level is then measured from the outer and flat points of a very extended radial profile of the core galaxies using the \textit{radial distance indicator} procedure\footnote{In a system where there is a clear BCG we would directly measure the sky level by fitting elliptical isophotes to the BCG and extending the profile to the radius where the profile flatters due to the sky emission.} developed in \citet[][]{Montes2014} and also used in \citet[][]{Montes2018, Montes2019}.

We create a map where the distance to each pixel on the images is computed as the elliptical distance to its nearest core galaxy. The shape parameters of the three core galaxies needed to build the ellipses were obtained by fitting elliptical isophotes to each of them separately, using the \verb|isophote| package \citep[using the iterative method described by][]{Jedrzejewski1987} in \texttt{Photutils} \citep{Bradley2020}. The fitting is performed to where the flux contribution starts to be dominated by that from the other two neighbouring galaxies. In order to estimate an accurate central position of each core galaxy, we calculate the ellipse fitting twice: a first run leaving all the parameters free; then a second in which we fixed the centre to the median fitted central coordinates returned by the previous iteration. From the distances map, we extract logarithmically spaced annuli that are then used to extract the radial profiles in the images. The shape of the final annuli apertures are shown in the bottom panel of Figure~\ref{fig:GroupIm}. Prior to the extraction of any profile, we built a mask with all the sources other than the core galaxies masked (see details on the masking process in Section~\ref{subsec:mask}).

By applying the above procedure over the individual $g$, $r$, and $i$ bands, we obtained extended radial profiles up to $\sim$170~arcsec ($\sim$580~kpc) for our sky subtraction determination. For each annulus of those very extended profiles, the sky flux was obtained by averaging the pixel flux values by applying a 3$\sigma$ clipping rejection algorithm. The sky level in a given band is defined as the median value of the radial profile points located in the outermost and flattest part of the profile (i.e. at radii $>350$~kpc). We subtracted from the whole images in each band the following sky values in counts: $\rm{sky}_g$=~--6.3~$\times$~10$^{-4}$, $\rm{sky}_r$=~--1.72~$\times$~10$^{-3}$, $\rm{sky}_i$=~--9~$\times$~10$^{-5}$.

\subsection{Modelling and removal of the scattered light} \label{subsec:scatlight}

The light coming from an astronomical source is scattered while passing through every optical element such as the atmosphere, the telescope, the instrument, and the detector. The point spread function (PSF) models and characterises how the light of a point source is scattered. The extended wings of the PSF of bright foreground stars can add an artificial flux component to the actual amount of light of the object \citep[e.g.][]{Uson1991, Michard2002, Slater2009, Sandin2015, TrujilloFliri2016, InfanteSainz2020}. As a consequence, it is important to construct a very extended PSF that reaches the whole extent of those wings, at least 1.5 times the size of the source of interest \citep{Sandin2014}. However, previous works have found that not only the light from bright stars affects measurements of LSB structures. Properties of the faint outskirts of galaxies or galactic discs such as thick discs or stellar haloes can be severely overestimated due to the effect of the PSF on the extended sources \citep[e.g.][]{Zibetti2004a, deJong2008, Sandin2014, TrujilloFliri2016, Peters2017, Comeron2017, MartinezLombilla2019b}. This is particularly important for small or compact sources like the ones we are studying in this work. For these reasons, it is crucial to model the scattered light from bright stars in the field as detailed in Section~\ref{subsubsec:brighStars}, and also to characterise how the very extended wings of the PSF affect extended objects such as the galaxies of our group, as we describe in Section~\ref{subsubsec:2Dmodels}.

In the following subsections we explain the characteristics of the very extended HSC PSFs used in this work and how we modelled and removed the scattered light from both the bright stars and the core galaxies of our group.

\subsubsection{Point-spread function (PSF)} \label{subsubsec:PSF}

In this work, we use our own custom-made extended PSFs, built from the wide layer HSC-SSP PDR2 data, for each of the three $g$, $r$, and $i$ bands. A careful measurement of such PSFs is essential to remove the scattered light from the images, without which the deep data cannot be exploited to its maximum.

Our HSC-SSP PDR2 PSFs were built following the method\footnote{See the detailed procedure of how to build an extended PSF in: \url{https://www.gnu.org/software/gnuastro/manual/html_node/Building-the-extended-PSF.html}} outlined by \citet{InfanteSainz2020}, which was designed for the Sloan Digital Sky Survey (SDSS). This method is based on the sampling of 425 point-like sources in each band, at different brightness levels, to estimate an average extended PSF of the whole data set by median-clipping star stacking. Three different ranges of brightnesses are chosen to build the so-called inner, intermediate and outer parts of such PSFs. Faint and non-saturated stars are used to reconstruct the PSF core, since the cores of bright stars are highly degraded due to the CCD pixel saturation. Because our PSFs are only used to remove scattered light from saturated sources, brighter than $m_{g}$=14~mag, the inner core and intermediate part do not play a significant role in the subtraction step, described in Section~\ref{subsubsec:brighStars}.

Bright stars are used to derive the faint PSF wings at their maximum available extension. For the SDSS survey, \citet{InfanteSainz2020} were able to yield a PSF radius of 8~arcmin by choosing around $1000$ sources with magnitudes below $7$. However, SDSS covers 10 times more sky area with respect to HSC. This difference reduces the total number of available sources and, consequently, limits the maximum reachable radial extension of the PSFs. However, although the vast majority of the brightest stars lie above $7$ magnitudes in the available field of view (FOV) of the HSC survey, we were able to select a large enough number of sources with magnitudes between $4$ and $8$ in each band to build reliable extended PSFs. 

Multiple quality checks were applied along the process in order to remove external light contamination caused by spurious sources in the vicinity of the selected stars used to create the PSFs. First, we rejected those stars which have very bright companions according to the available stellar catalogues. Second, we detected and masked objects and structures that contaminate the FOV using GNU Astro Astronomy Utilities (Gnuastro)\footnote{Webpage of Gnuastro: \url{https://www.gnu.org/software/gnuastro}} \verb|NoiseChisel| and \verb|Segment| \citep{Akhlaghi2015, Akhlaghi2019}, as performed in \citet{InfanteSainz2020}. These tools were also used to estimate the sky noise level accurately, measuring it from the non-masked regions. Third, we removed those stars that still exhibited external light contamination from non-detected sources or ghosts. The final number of bright stars that could be used were around $425$ in each band, which allowed us to reach a signal to noise ratio (S/N) above $3$ along the entire radial profile of the final PSFs.    

\begin{figure}
\begin{center}
\includegraphics[width=\columnwidth]{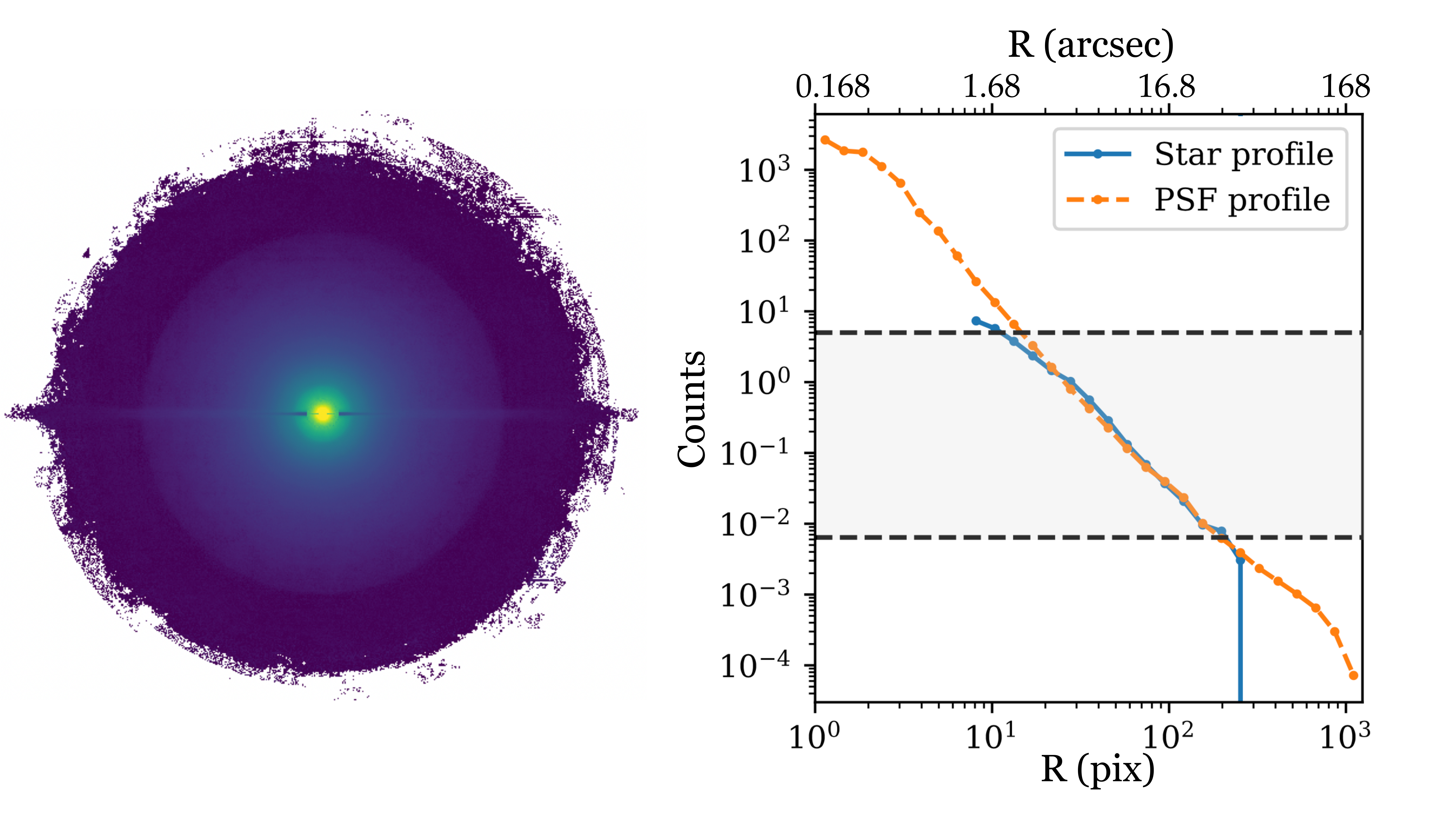}
\caption{\textit{Left}: Deep and very extended HSC-PDR2 PSF image in $i$ band. The PSF was obtained by combining the PSF of bright saturated stars for the outer part, intermediate-magnitude stars for the intermediate transition region, and high-S/N faint stars for the inner part. This PSF extends to a radius $\rm{R} \sim 1607$~pix (4.5~arcmin). \textit{Right}: Profiles in the $i$ band of the fitted bright star S3 (blue) and of the flux scaled model HSC PSF (orange). The horizontal grey area between the black dashed lines indicates the radial region selected for the profile fitting between the star and the model PSF for this particular case. 
  \label{fig:PSF}}
\end{center}
\end{figure}

Finally, intermediate-magnitude point-like sources are necessary to join together the inner-core ($4-5$~arcsec) and outer-wings ($20-25$~arcsec) parts. The $i$ band PSF is shown in Figure \ref{fig:PSF}, where it can be seen that the PSF extends to a radius of 1607~pix, which is equivalent to $4.5$~arcmin. For the $g$ and $r$ bands, the PSF extends to similar radii. The blank pixels in Figure~\ref{fig:PSF} are due to insufficient data at these locations, whereas the central elongated artefact along the horizontal axis is due to the saturation of the brightest stars used to build the outer part.    


\subsubsection{Bright star subtraction} \label{subsubsec:brighStars}

We subtracted the scattered light contamination from the brightest stars in the field by following procedures similar to those presented in \citet{Roman2020, InfanteSainz2020} and \citet{Montes2021a}. In this method, the flux of the selected stars is fitted in each band and then the flux of the appropriate band model HSC PSF is scaled to that of the star. In this image, we fitted the three stars in our FOV that are brighter than $m_{g}$=14~mag (see their locations in Fig.~\ref{fig:GroupIm} and parameters in Table~\ref{tab:brightStars}). 

We started by masking all the sources in the field other than these 3 stars by constructing a dedicated mask for each star (see details of the masking process in Section~\ref{subsec:mask}). The next step was to obtain an accurate position of the centre of each star. The \verb|centroid_1dg| package in Photutils \citep{Bradley2020} allows us to find the centre of these saturated stars by fitting 1D Gaussians to the marginal x and y flux distributions within a radius of a few arcseconds around their central regions.

Once the images are properly masked and the centres of the stars carefully determined, we are able to fit the flux. First, we fitted the flux from the stars. The order of the fits were based on the brightness, from brightest (i.e. star S1), to faintest (i.e. star S3). We derived radial profiles of the flux from the stars using logarithmically spaced circular annular apertures (using \verb|CircularAnnulus| task in Photutils). The radial profile is composed of the average flux of each annulus obtained after applying a 3$\sigma$ clipping rejection algorithm to minimise contamination from residuals or undesirable artefacts in the data. We then extracted the radial profiles of the model HSC PSFs following the same procedure.

The next step is the flux calibration. This is, to scale the radial profiles of the model HSC PSFs to those of the stars in each band. We selected a range in radius to obtain the scale factor considering that our main goal is to characterise the outer extended wings of the PSF of the stars to minimise the contamination in the outer parts, where the group light could be affected. The radial range is then from 0.05 times the saturation level of the images to 3 times their background value. The scale factor is the ratio between the profile of the stars and the corresponding model HSC PSF within the selected radial range. To ensure that we measured a robust scale factor, we minimised the fitting by optimising the values of the scale factor at each annulus in the selected radial range of the profiles. We scaled each model HSC PSF profile by the corresponding scale factor. 

Lastly, we subtracted the flux-calibrated model HSC PSF of the three stars all at once. To do that, we scaled the model HSC PSF images using the above calculated scale factor for each of the stars. Then, we put those individual images of the scaled PSFs in a new blank image the size of the FOV at the position of the corresponding star. This final image is the scattered field image, and is what we subtracted from the data to obtain a FOV corrected for the effect of the PSF in bright stars. We repeat this process for each $g$, $r$, and $i$ band image. 

Figure~\ref{fig:GroupIm} shows the image of our FOV. The original image is seen in the top panel, while the bottom row shows the same region with the bright stars subtracted and masked for each band separately (the core galaxies were also modelled and subtracted as explained in Section~\ref{subsubsec:2Dmodels}). Some residual are left after modelling bright stars, especially in the inner regions affected by internal reflections \citep{InfanteSainz2020}.



\begin{table}
	\centering
	\caption{Main parameters of the three bright stars to which the PSFs were scaled and subtracted (Section~\ref{subsubsec:brighStars}). The columns indicate the star name (S1, S2, and S3 as indicated in Figure~\ref{fig:GroupIm}), the star coordinates, its corresponding Gaia ID from the Data Release 2 \citep[DR2;][]{Collaboration2018}, and the apparent magnitude in $G$ band obtained from the HSC-PDR2 bright masks data \citep{Coupon2018}.  }
	\label{tab:brightStars}
    \begin{tabular}{lcccc}
        \hline
         Star & R.A. (J2000) & DEC (J2000) & Gaia ID & $m_{G}$ \\
           & [h:m:s] & [d:m:s] & DR2 & [mag] \\
            \hline
            S1 & 2:23:14.9441 & -5:28:37.713 & 2488659911371254528 & 12.47 \\
            S2 & 2:23:16.4314 & -5:28:58.823 & 2488659911371254784 & 10.41 \\
            S3 & 2:23:31.0018 & -5:25:56.160 & 2488660598566021376 & 13.80 \\
        \hline
    \end{tabular}
\end{table}

\subsubsection{2D PSF-deconvolved models of the galaxies in the group} \label{subsubsec:2Dmodels}

The scattered light from the galaxies situated in the core of the group may also contribute to overestimating the total amount of IGL in the system and could affect the measured colour distribution of this faint light. To avoid these issues, we built 2D models deconvolved with our extended PSF of the 3 galaxies in the group core: BCG, 1660730 and 1660646. These models are obtained from images with the sky and bright stars subtracted.

We generate the 2D models using \textsc{imfit}\footnote{Precompiled binaries, documentation, and full source code (released under the GNU Public License) are available in \url{https://www.mpe.mpg.de/~erwin/code/imfit/} } \citep{Erwin2015}. During the fitting, \textsc{imfit} allows for the convolution of analytical functions with the image of the extended PSF (see Section~\ref{subsubsec:PSF}). This technique allows the creation of a 2D PSF-convolved model of a source and has been widely used in previous studies \citep[e.g.][]{TrujilloFliri2016, Peters2017, MartinezLombilla2019b}.

\begin{figure*}
\begin{center}
\includegraphics[width=\textwidth]{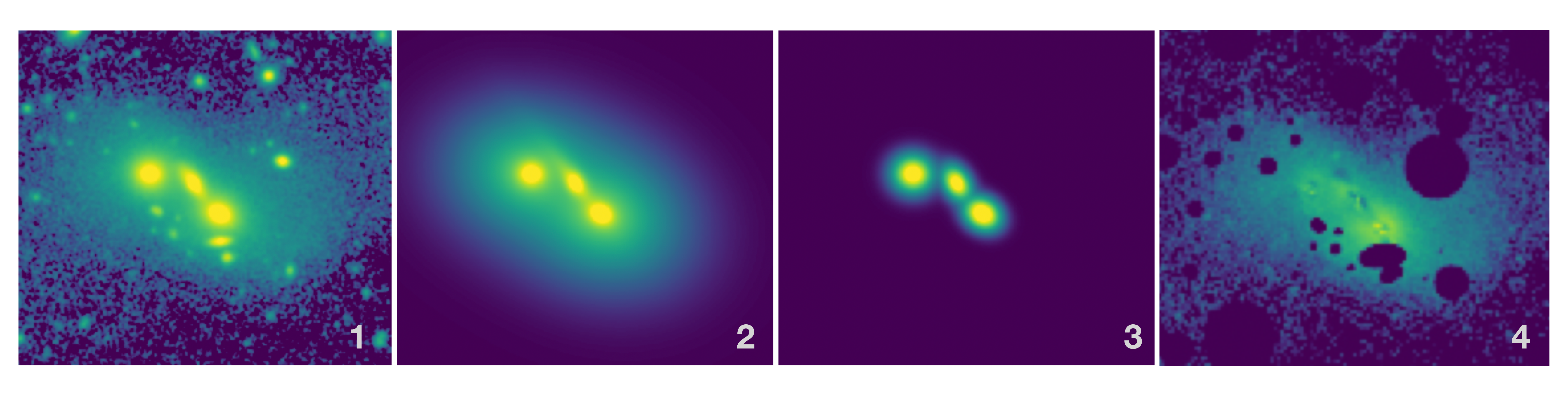}
\caption{Images of a zoom-in region of $\sim 200 \times 200$~kpc over the core region of the group showing the steps of the process for modelling and subtracting the 2D PSF-convolved models of the three core group galaxies. 1) HSC PDR2 data in $i$ band, sky and star-subtracted; 2) 2D PSF-convolved models of the three core galaxies and the IGL; 3) 2D PSF-convolved models of the three core galaxies only; 4) HSC PDR2 data in $i$ band, sky, star and galaxy-subtracted after removing the 2D PSF-convolved models of the three core galaxies shown in image 3.
  \label{fig:2Dgalmod}}
\end{center}
\end{figure*}

In order to model the BCG, 1660730, and 1660646, we used a S{\'e}rsic function \citep{Sersic1968} to reproduce the galaxies and an exponential function for the diffuse component. 1660730 needed an additional exponential function to model a disk-like structure in this galaxy. The details on the parameters and values used for each function are given in Table~\ref{tab:imfitparam} and the images of the models are shown in Figure~\ref{fig:2Dgalmod}. For the initial conditions we used the outcomes from the elliptical isophote fitting performed to extract the sky background value (Section~\ref{subsec:skySBP}).

We ran \textsc{imfit} twice for each galaxy in each band (6 times in total). First, we fit the light from the galaxies and the diffuse component accounting for the PSF effects. In the second fit, we take the best-fit parameters of the best fitting of the galaxies --this is, excluding the exponential component of the diffuse light-- and ran \textsc{imfit} without any minimisation. This gives the 2D PSF-convolved models of the galaxies in the group core.

Finally, we subtracted the image with the 2D PSF-convolved models of the three core galaxies from the original image with the bright stars already subtracted. The resulting image only contains the diffuse light component, corrected for any scattered light from the galaxies, and a residual fraction of light the analytical galaxy models are not able to replicate such as intrinsic asymmetries in the morphology of the observed galaxies (hereafter, we refer to this component as the residuals). By repeating this process for the three $g$, $r$, and $i$ bands, we produce the final images of the diffuse light that we use for the rest of this analysis. Figure~\ref{fig:2Dgalmod} shows the images at each step of the process and the bottom panels in Figure~\ref{fig:GroupIm} show the images of the IGL in each band.

\begin{table*}
	\centering
	\caption{Best fit parameters values obtained with \textsc{imfit} for each function required to model the three core group system of galaxies and IGL in the $g$, $r$, and $i$ bands as described in Section~\ref{subsubsec:2Dmodels}. The S{\'e}rsic functions model the galaxies (or the spheroidal core in the case of 1660646) defined by four main parameters: the position angle (PA) in degrees; the ellipticity ($\varepsilon$); the S{\'e}rsic index ($n$); and the effective radius ($r_{\rm{e}}$) in arcsec. The exponential function models the faint IGL structure (and the disc component in 1660646) and is characterised by three parameters: the position angle (PA) in degrees; the ellipticity ($\varepsilon$); and the exponential scale length ($h$) in arcsec. }
	\label{tab:imfitparam}
    \begin{tabular}{cccccccccccc}
        \hline
          Params & \multicolumn{3}{c}{BCG} & & \multicolumn{3}{c}{1660646} & & \multicolumn{3}{c}{1660730} \\
            \hline
            S{\'e}rsic & $g$ & $r$ & $i$ & & $g$ & $r$ & $i$ & & $g$ & $r$ & $i$ \\
            \hline
                PA [deg] & $63 \pm 3$ & $60 \pm 2$ & $58 \pm 1$ & & $31 \pm 3$  & $29 \pm 3$  & $28 \pm 4$  & & $267 \pm 4$ & $264 \pm 3$ & $272 \pm $ 6\\
                $\varepsilon$ & $0.36 \pm 0.06$ & $0.30 \pm 0.03$ & $0.28 \pm 0.02$ & & $0.16 \pm 0.05$ & $0.14 \pm 0.06$ & $0.10 \pm 0.06$ & & $0.20 \pm 0.04$ & $0.26 \pm 0.06$ & $0.14 \pm $ 0.03\\
                $n$ & $1.46 \pm 0.19$ & $2.03 \pm 0.30$ & $2.19 \pm 0.17$ & & $1.53 \pm 0.19$ & $1.4 \pm 0.55$ & $0.89 \pm 0.37$ & & $1.45 \pm 0.15$ & $2.22 \pm 0.43$ & $2.79 \pm 0.34$ \\
                $r_{\rm{e}}$ [arcsec] & $0.79 \pm 0.03$ & $1.00 \pm 0.01$ & $1.07 \pm 0.07$ & & $1.06 \pm 0.09$ & $0.44 \pm 0.03$ & $0.30 \pm 0.05$ & & $0.78 \pm 0.03$ & $1.13 \pm 0.02$ & $1.16 \pm 0.16$ \\
            \hline
            Exp. (IGL) & & & & & & & & & & & \\
            \hline
                PA [deg] & $100 \pm 8$ & $109 \pm 11$ & $93 \pm 9$ & & $166 \pm 15$ & $36 \pm 16$ & $46 \pm 7$ & & $24 \pm 8$ & $80 \pm 15$ & $73 \pm 14$ \\
                $\varepsilon$ & $0.21 \pm 0.06$ & $0.29 \pm 0.05$ & $0.48 \pm 0.12$ & & $0.55 \pm 0.13$ & $0.89 \pm 0.11$ & $0.74 \pm 0.12$ & & $0.50 \pm 0.08$ & $0.32 \pm 0.11$ & $0.53 \pm 0.18$ \\
                $h$ [arcsec] & $4.69 \pm 1.03$ & $5.04 \pm 0.89$& $9.39 \pm 2.26$ & & $3.72 \pm 1.01$ & $3.02 \pm 0.99$ & $7.59 \pm 1.42$ & & $4.88 \pm 1.05$ & $4.37 \pm 0.96$ & $8.16 \pm 1.65$ \\
            \hline
            Exp. (disc)  & & & & & & & & & & & \\
            \hline
                PA [deg] & - & - & - & & $36 \pm 2$ & $35 \pm 3$ & $35 \pm 2$ & & - & - & -  \\
                $\varepsilon$ & - & - & - & & $0.67 \pm 0.03$ & $0.66 \pm 0.02$ & $0.62 \pm 0.03$ & & - & - & -  \\
                $h$ [arcsec] & - & - & - & & $1.23 \pm 0.38$ & $0.74 \pm 0.21$ & $1.12 \pm 0.13$ & & - & - & -  \\
            \hline
    \end{tabular}
\end{table*}

\subsection{Dedicated semi-automatic mask} \label{subsec:mask}

Masks play a key role in the analysis of LSB structures because any undesirable source of light, even from very faint objects, can contribute to incorrect flux measurements. The aim of masking is to block the light from all fore- and background objects in the field of view except for the target. However, the combination of faint and extended structures that are present in LSB studies makes the masking process very complex and highly human-dependent. Source detection and deblending algorithms usually struggle to separate faint extended objects --particularly from the background-- with the precision required here. In this Section, we describe how we have developed a semi-automatic Python-based process, optimised to build the very extended and conservative masks required for the analysis of the IGL in our group of galaxies (see bottom panels of Figure~\ref{fig:GroupIm}). This method is built from modular and object-oriented Python scripts, that allow for an easy implementation of modifications and improvements. We used Photutils \citep{Bradley2020} for the image segmentation using \verb|detect_threshold| and \verb|detect_sources| to detect astronomical sources, and \verb|deblend_sources| to deblend the detected sources. In all the cases, the threshold for detection was 1.1~$\sigma$ above the background level.

In general, we generated our masks in three steps: 1) We first build a mask optimised to detect and deblend bright extended objects --i.e. a cold mask--, for example the bright stars and the galaxies in the group; 2) On top of that, we made a mask optimised for faint small objects --i.e. a hot mask--, such as background galaxies or foreground dwarf galaxies; And then 3) by visual inspection and using an interactive interface provided by Python, we mask any undetected source or structure in the resulting combined hot+cold mask through an iterative process that allows us to improve the masking with every iteration. 

The ``hot+cold'' method has been previously used in deep images where there are a wide variety of objects in the field in terms of flux and radial extension \citep[e.g.][]{Rix2004, Montes2018, Montes2021a}. This is because a single source detection algorithm is not feasible in those cases as it cannot simultaneously detect such a wide variety of sources robustly. Another situation where these algorithms have problems is when background sources --especially faint ones-- overlap bright objects like the galaxies of our group. To address this issue, we unsharp-masked the original image prior to the application of the hot mask to increase the contrast. To do that, we convolved the image with a Gaussian filter \citep[\texttt{Gaussian2DKernel} from Astropy;][]{Astropy2018} with $\sigma=$5~pixels, and subtracted it from the original. 

We followed a conservative approach throughout the masking process. We applied the whole ``hot+cold'' method over the deep $gri$ stacked image as the stacking technique enhances the sources' brightness allowing for more extended masks. To completely cover the faint outskirts of the sources, we expanded each of the hot and cold masks by convolving them with 2D Gaussian filters with $\sigma =$~1~pixels and $\sigma =$~2~pixels respectively. We also used the large bright star masks (brighter than $g$= 18~mag) provided by the HSC-PDR2 \citep[][]{Coupon2018, Aihara2019} as an input for our cold mask script. To account for detector issues we applied the \verb|SENSOR_EDGE|, \verb|CROSSTALK|, and \verb|BAD| mask planes provided in the HSC-PDR2 coadds \citep[][]{Bosch2018, Aihara2019}. These detector issues are the reason why some rows and columns are completely masked in the images showed in the bottom panels of Figure~\ref{fig:GroupIm}. All these techniques together ensure the total masking of any unwanted light. 


Sometimes the masking process required us to unmask certain sources. To do that, we visually determined a region around the target using the software SAOImageDS9 \citep{Joye2003}. The region is then incorporated to the Python scripts using the \verb|read_ds9| function from the \verb|regions| package in Astropy. This is the most human-dependent step of the method but it can be done before starting the masking process.  

In a complex system such as a group of galaxies, we needed to build different masks depending on the analysis step and the source of interest. Based on the general ``hot+cold'' method explained above, we constructed a set of 9 different masks. In all cases, we used the position on the sky provided by the GAMA catalogues to identify the group members. The specific details of each of these masks, as well as the relevant analysis step they refer to, are listed here:

\begin{itemize}
    \item \textit{Bright star subtraction masks} (Section~\ref{subsubsec:brighStars}): To subtract the bright stars in the field we needed to mask all the sources except for our targeted stars S1, S2, and S3 (see star locations in the top panel of Figure~\ref{fig:GroupIm}). This is a multi-step process combined with fitting the flux of the stars. We first masked all the sources within the FOV and added an elliptical mask centred on the BCG to completely cover the faint light from the core region (ellipse with a semi-major axis of $\rm{SMA}=105$~arcsec, and a semi-minor axis of $\rm{sma}=58$~arcsec). At this stage we had all the sources masked. From here, we built a mask for each star taking into account that stars S1 and S2 are very close to one other. As the flux fitting started with the brightest star, we first unmasked S1 by isolating its flux from that of S2 as much as possible while still allowing for a reliable fitting. The unsharped mask technique is applied here to properly mask the background sources overlapping S1. Once the S1 mask was ready, we performed the flux fitting and subtraction of the star as explained in Section~\ref{subsubsec:brighStars}. We then started the mask for S2. First masking the residuals derived from the subtraction of S1. Then we unmasked S2 and added a mask to the overlapping background sources. We proceeded in the same way as before with the fitting and subtraction of S2. Finally, we built the mask for S3 by repeating the same process. In all cases, the horizontal spikes produced by the saturated/bleeded pixels were also masked. The very last step was to mask the residuals of S3. Thereby, the mask of the star-subtracted image was produced.  
    
    \item \textit{Group core mask}: (Sections~\ref{subsec:skySBP}, \ref{subsubsec:2Dmodels}, \ref{subsec:IGL_SBP}, and \ref{subsec:IGL_color}): From the mask of the star-subtracted image, we unmasked the core region of the group --this is, removed the large elliptical mask centred on the BCG--, and then masked every source within that area except the core galaxies (the BCG, 1660730 and 1660646). We also left unmasked the two foreground galaxies below the BCG (F1 and F2 in Figure~\ref{fig:GroupIm}) as their flux might affect that of the BCG. To remove any influence of those galaxies, we first performed an elliptical isophote fitting of their flux allowing for a 3$\sigma$ clipping algorithm. The result of the fitting was used to built a 2D model with \verb|build_ellipse_model| in Photutils. Finally, the 2D model was subtracted from the data. By masking the residuals derived form the subtraction of the galaxies F1 and F2, we obtained the group core mask. This mask is widely used in the analysis process as this is required to perform the second-order sky subtraction, is the input file to \textsc{imfit} to generate the 2D PSF-deconvolved models of the three galaxies in the core of the group, and is the mask for extracting the surface brightness and colour profiles of the core group members.    
    
    \item \textit{BCG mask} (Section~\ref{subsubsec:IGL_frac_2Dmod}): Following the above procedure for the foreground galaxies, we took the \textit{group core mask} and added a mask for 1660730 and 1660646. Thus, in this case every source is masked except for the BCG.  

    \item \textit{Diffuse light mask} (Sections~\ref{subsubsec:2Dmodels}, \ref{subsec:IGL_SBP}, \ref{subsec:IGL_color}, and \ref{subsubsec:IGL_frac_2Dmod}): Once the three galaxies in the core of our group were modelled and subtracted from the data, we also masked over the central residuals remaining after those subtractions. We called this the diffuse light mask as it only allows for the flux from the diffuse light component of the system.   
    
    \item \textit{Group mask} (Sections~\ref{subsec:IGL_fraction} and \ref{subsubsec:IGL_frac_SBcut}): A mask that covers every source but the light from the group members is necessary to obtain an estimation of the total group light in order to determine the IGL fraction of that light. We unmasked the remaining group members from the group core mask and masked any undesirable background and foreground source that falls over the group galaxies light by applying the unsharped mask technique.     
    
    \item \textit{RMS mask} (Section~\ref{subsec:sblim}): In this case we needed to completely mask all the sources and the faint light. To do that, we took the core mask and added the elliptical mask used in the bright stars subtraction mask to it, to cover the group core and the IGL ($\rm{SMA}=105$~arcsec and $\rm{sma}=58$~arcsec).   

\end{itemize}

\subsection{Surface brightness limits} \label{subsec:sblim}

Any LSB study requires an estimation of the surface brightness limit. This limit sets an initial framework for the analysis in terms of the depth of the data, which constrains the potential detectable structures. We determined the surface brightness limit for each of the three $g$, $r$, and $i$ bands independently following the procedure of Appendix A in \citet{Roman2020}. First, we had to calculate the root-mean-square (RMS, i.e. a measurement of the scatter) of the data by randomly placing 2500 boxes of 10~$\times$~10~arcsec$^{2}$ ($\sim$34~$\times$~34~kpc$^{2}$) throughout each image masked with the RMS mask (see Section~\ref{subsec:mask}).  

The surface brightness limits of our data at $3 \sigma$ within $10 \times 10$~arcsec$^{2}$ boxes are as follows: $\mu _{g} ^{\rm{lim}}=30.76$~mag~arcsec$^{-2}$; $\mu _{r} ^{\rm{lim}}=29.82$~mag~arcsec$^{-2}$; $\mu _{i} ^{\rm{lim}}=29.41$~mag~arcsec$^{-2}$.


\section{Results} \label{sec:results}

At this stage, the images are sky subtracted and the bright stars and core galaxies are modelled and subtracted to account for the influence of the extended wings of their PSF. Therefore, our HSC-PDR2 ultra deep images are ready for the analysis of the IGL in the group 400138.

In this section we show the results of the study of the diffuse light within our group of galaxies. We detect and separate the IGL component from other structures such as the halo of the BCG. To do that, we analyse the distribution of that diffuse light in Sections~\ref{subsec:IGL_image} and \ref{subsec:IGL_SBP}, its stellar populations using the colour information in Section~\ref{subsec:IGL_color}, and the amount of IGL in the group in comparison with the total light of the system in Section~\ref{subsec:IGL_fraction}.

\begin{figure*}
\begin{center}
\includegraphics[width=14cm]{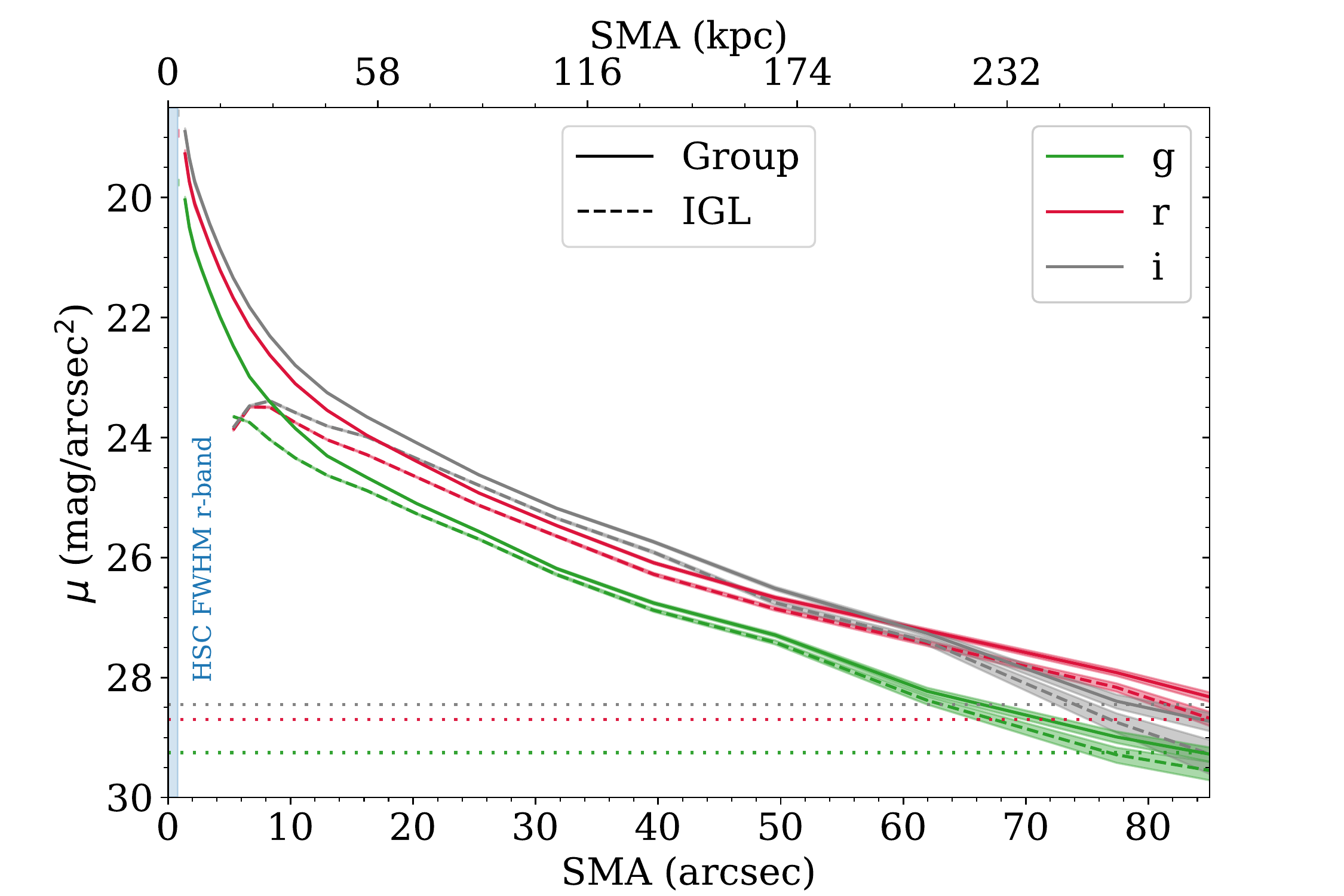}
\caption{Surface brightness profiles at each $g$ (green), $r$ (red), and $i$ band (grey) of the core group galaxies (solid lines) and the IGL component (dashed lines). These profiles are corrected for Galactic extinction, surface brightness dimming, and k-corrected. The horizontal dotted lines are the surface brightness limits of each corresponding band from Section~\ref{subsec:sblim}, also corrected for surface brightness dimming and k-correction (colour-coded as the profiles). The vertical blue region represents the HSC FWHM of 0.7~arcsec ($\sim$4.2~pix) obtained from point sources in $r$ band. The IGL of the 400138 galaxy group shows a smooth exponential decay, brighter in redder wavelengths, that extends from SMA$>55$~kpc up to SMA$\sim$275~kpc in all bands with reliable detection below 26~mag~arcsec$^{-2}$ down to the data surface brightness limits.
  \label{fig:IGLsbp}}
\end{center}
\end{figure*}

\subsection{Diffuse light image and distribution} \label{subsec:IGL_image}

In this Section we briefly describe the visual properties of the diffuse light component. The main result is that we have indeed detected the diffuse component, including the IGL, in each of our three bands. To observe and obtain reliable measurements of the IGL is a technical challenge due to its faint nature. At those low surface brightness levels ($\mu _{g}>$~26~mag~arcsec$^{-2}$), even the faintest source of light could affect what we consider as IGL emission (see Section~\ref{sec:prepdata} for details about the light contributions that have been removed from the images). One of the main sources of uncertainty in this case would be the effect of the PSF in the light profiles of the galaxies, producing a fake stellar halo. However, as we are still detecting diffuse light \emph{after} deconvolving with the PSF, we can confidently claim that our IGL measurements are not overestimated by the contribution of the scattered light of the core group galaxies. In the same way, the flux from bright stars within the field, that may produce a faint light component over the whole group, has also been effectively removed. This is the first IGL detection in a group of galaxies in the HSC-PDR2 ultra deep images. Further IGL analysis could be carried out in a more representative sample of galaxy groups (see Section~\ref{subsec:FurtherAn_Discus}). 

The bottom panels in Figure~\ref{fig:GroupIm} show the IGL radial distribution. The IGL emission is more intense at smaller radii in the three bands. The inner IGL distribution follows the ellipticity of the system formed by the three core galaxies (the BCG, 1660730, and 1660646), with a position angle orientated towards the North-East. However, the diffuse light also extends towards the North-West (upper right direction in Figure~\ref{fig:GroupIm}) in the outermost parts showing a tail-like structure. The tail reaches about 91~arcsec from the BCG and is around 22~arcsec wide in $i$ and $r$ bands ($\sim$314~$\times$~75~kpc$^{2}$), being fainter and less extended in $g$ band ($\sim$265~$\times$~66~kpc$^{2}$). Looking at the location of the galaxy group members in Figure~\ref{fig:GroupIm}, it is possible that the gravitational potential of the galaxy 1660615 is causing this additional light extension over the main central component. The diffuse light does not extend to the furthest group members 2301069 and 1660545 at the surface brightness limit of our data. As stated in Section~\ref{subsec:cirrus}, small and compact Galactic cirrus structures are a potential source of uncertainty because they might be observed as faint regions in optical images, resembling low surface brightness structures. Far-infrared detectors are blind to these cirrus due to their poor angular resolution meaning that the extended diffuse structure that extends towards the North-West from the core of 400138 could be a cirrus feature. However, the association with confirmed member 1660615 supports the IGL nature of this faint tail.

\subsection{IGL radial surface brightness profile} \label{subsec:IGL_SBP}

To study how the IGL is distributed radially, we extracted identical radial surface brightness profiles of the star-subtracted core members of the group (i.e., the image with the core galaxies), and the diffuse light component from the 2D star- and galaxy-subtracted images (i.e., the image without the core galaxies). We derived those 1D profiles in each of the three $g$, $r$, and $i$ bands separately. To do that, we used the \textit{radial distance indicator} procedure described in Section~\ref{subsec:skySBP}.  

In this \textit{radial distance indicator} approach we consider the centre of mass of the 3 core galaxies (the most massive members) as the centre of the system. This is because the BCG is not located at the centre of the IGL distribution (see Figure~\ref{fig:GroupIm}). Elliptical apertures centred in the BCG would cause larger sources of uncertainties in the IGL measurements from pixels that do not belong to the IGL but that would be in the elliptical annulus bins. Another possibility would be to choose the centre of mass of the group but it is located at $\sim 14$~arcsec from the BCG in the North-West direction, so the offset from the centre of IGL distribution is even larger that that of the BCG. On the contrary, the distance annulus apertures used in this work follow the visible IGL distribution as shown in Figure~\ref{fig:GroupIm}, providing confidence in the good agreement between the centre we used and the centre of mass of the IGL. We measured the centre of mass (or luminosity) of the images with only the IGL emission (this is, the 2D star- and galaxy-subtracted images). We obtained that the IGL centroid should be within $R<3$~arcsec from the centre we used. This potential small offset would imply only a minimal difference in the measurements of the light of the three core galaxies or the very inner regions of the IGL (SMA~$>15$~arcsec).

From the annuli apertures, we built 21 radial bins logarithmically spaced around the centre of mass of the system composed by the three core galaxies, up to a SMA of 85~arcsec (290~kpc) in the three $g$, $r$ and $i$ bands. The value of each radial point of the profile is the median surface brightness of the corresponding annulus obtained after performing a 3$\sigma$ clipping rejection algorithm. All the surface brightness profiles (i.e. the whole profiles) are corrected for Galactic absorption ($A(g)=0.091$, $A(r)=0.066$, and $A(i)=0.050$~mag) and are $K$-corrected ($K$-corr$(g)$=0.68, $K$-corr$(r)$=0.29, and $K\rm{-corr}(i)$=0.13~mag). We used the values provided for the BCG in the GAMA catalogues: the Galactic absorptions from \citet[][]{Baldry2018}, and the $K$-corrections from \citet{Loveday2012}. As the group 400138 is at redshift $z\sim0.21$, we also accounted for the surface brightness dimming in the form $(1 + z)^{-4}$ \citep[][]{Tolman1930, Tolman1934}. The error of each surface brightness value in the profile is estimated as in \citet[][]{MartinezLombilla2019a}. This is the combination of the Poisson error of the annulus (associated with the signal of the source in each ellipse), and the uncertainty on the estimation of the sky due to fluctuations within the area of each ellipse (associated with the sky RMS obtained in Section~\ref{subsec:sblim}).

Figure~\ref{fig:IGLsbp} shows the resulting surface brightness profile of the core galaxies of the group, and the diffuse component in each of the $g$, $r$, and $i$ bands. The shaded regions represent the uncertainties on the flux determination. Both the profiles of the group and of the IGL are brighter in $i$ and $r$ band than in $g$ band. This can also be appreciated in the bottom panels of Figure~\ref{fig:GroupIm}, where the diffuse light is less intense and covers a smaller area in $g$ band, while the higher S/N and more extended detection are in $r$ and $i$ bands. The IGL profiles show a flat shape in the inner regions (SMA $<10$~arcsec, $30$~kpc) due to the residual light from the haloes of the three galaxies not being perfectly modelled in the core region (see Section~\ref{subsubsec:2Dmodels}). In the case of the core of the group, the inner region accounts for the light from the three core galaxies. Then, at SMA~$>55$~kpc, there is a smooth exponential decay in brightness towards the outer parts tracing a faint light component, reliably detected down to the data surface brightness limits. This light can be considered as the IGL of the 400138 galaxy group. The profiles extend up to $\sim$~80~arcsec (SMA~$\sim 275$~kpc) in all bands, with S/N values $>1$ to this radial distance (see Figure~\ref{fig:GroupIm} for the shape of the apertures and the IGL extension over the image).

\begin{figure*}
\begin{center}
\includegraphics[width=14cm]{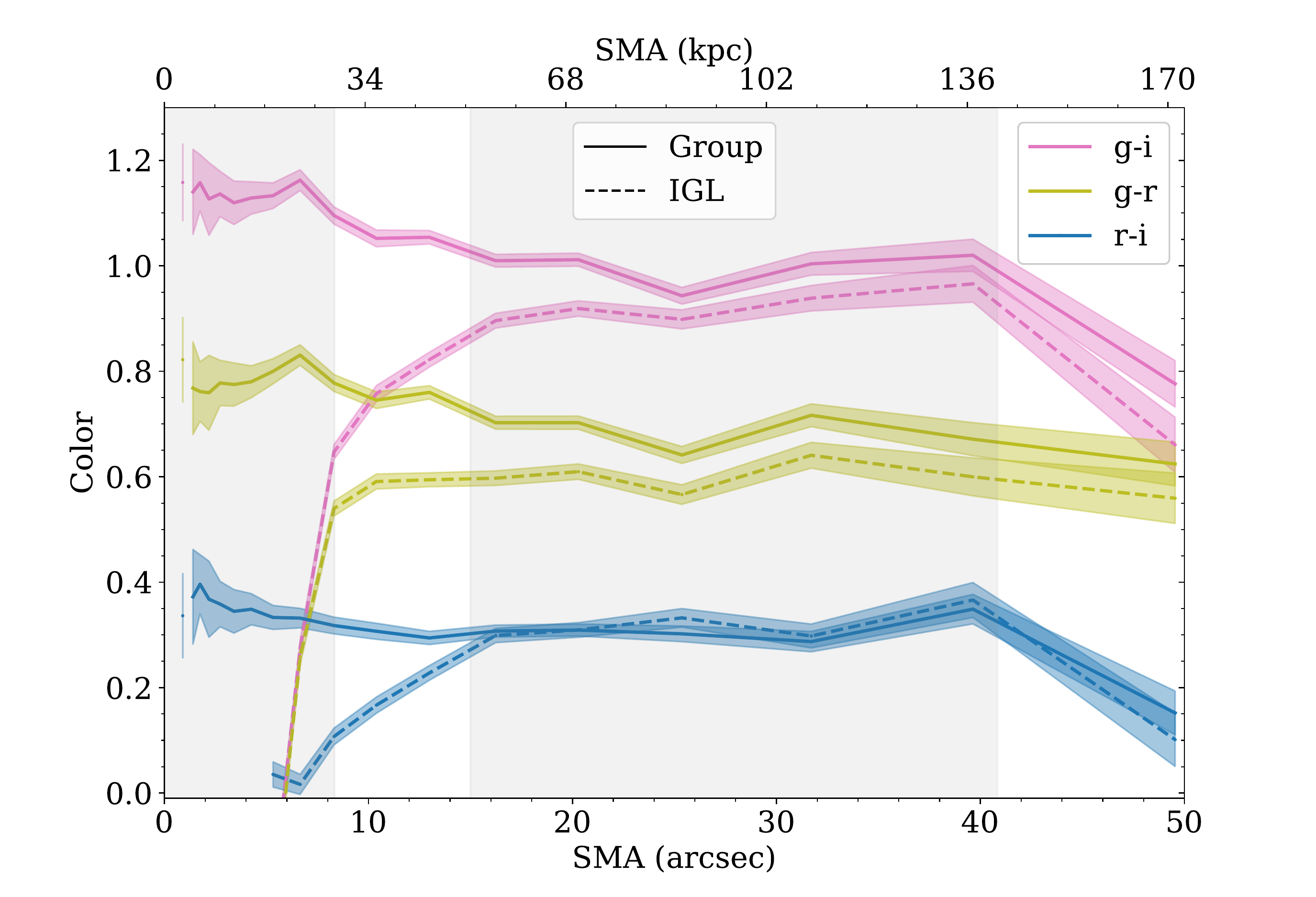}
\caption{($g-i$), ($g-r$), and ($r-i$) colour profiles in pink, yellow, and blue respectively. The solid lines represent the profiles extracted from the three core galaxies of the group, while the dotted lines represent those of the IGL component. The coloured shaded regions account for the profiles' associated errors. The grey regions indicate the range from which we obtain the mean colour values: SMA$<30$~kpc for the mean colours of the core galaxies of the group, and $50<$SMA$<140$~kpc for those of the IGL. The corresponding profile apertures at these distances are indicated in Figure~\ref{fig:GroupIm}. There is a gradient towards bluer colours in the transition region between the group and the IGL (SMA$<50$~kpc) indicating younger stellar populations in the IGL component. Beyond that ($50<$SMA$<140$~kpc) the IGL is the dominant component and the colour profiles show flat radial distributions within the errors.
\label{fig:IGLcolor}}
\end{center}
\end{figure*}

\subsection{IGL colour} \label{subsec:IGL_color}

Colour measurements can give invaluable information on the stellar populations that make up the IGL. Studying the radial colour distribution of the core galaxies and the IGL component, we can set age and metallicity limits on the stars making up the IGL, and also constrain the potential physical processes that built up this diffuse structure. We obtained three radial colour profiles of the diffuse light component for all our possible filter combinations: $g-i$, $g-r$, and $r-i$. The radial colour profiles were directly obtained from the subtraction of the surface brightness profiles in the corresponding bands in Section~\ref{subsec:IGL_SBP} and are therefore k-corrected and corrected for Galactic extinction. The uncertainties on the colour determination were obtained from the combination of the error values of the surface brightness profiles, providing comparable values to those in previous low surface brightness works using HSC-PDR2 data \citep[e.g.][]{Montes2021}.

Figure~\ref{fig:IGLcolor} shows these radial colour profiles out to 50~arcsec (170~kpc; see the example radial apertures over the IGL image in Figure~\ref{fig:GroupIm}). The solid lines indicate the colour profiles of the group members (galaxies + IGL), while the dashed lines are those with the core galaxies removed from the image, the IGL colour profiles. The radial colour profiles have four clear behaviours depending on the radial distance: 

- The inner part of the group members at SMA$<8$~arcsec (SMA$<30$~kpc), colour profiles are dominated by the emission from the galaxies. In the case of the IGL colour profiles, as those core galaxies were modelled and subtracted (see Section~\ref{subsubsec:2Dmodels}), the colour measurements are not reliable in that region. 

- Then, there is a transition region at $8<$SMA$<15$~arcsec ($30<$SMA$<50$~kpc) where the IGL gradually becomes less affected by the core members. At the same time, there is a clear colour gradient in the group profiles towards the blue. In this transition region, the pure IGL colour profiles (dashed lines) are still affected by the contribution from the residuals derived from the subtracted galaxy models as the separation between the galaxies and the IGL is inherently imperfect when using photometric data only \citep[see review by][and references therein]{Montes2022}. For that reason, this area does not provide reliable IGL colour measurements and is not used in this analysis. 

- In the intermediate part of the profiles at $15<$SMA$<41$~arcsec ($50<$SMA$<140$~kpc), both the IGL and group colour profiles show similar radial trends, almost flat within errors. At these distances, the IGL is the dominant component and shows bluer colours than that of the core galaxies of the group. 

- In the outer parts of the profiles at SMA$>41$~arcsec (SMA$>140$~kpc), we see a clear negative gradient in the $g-i$ and $r-i$ colours, while $g-r$ decreases slowly. This is mainly produced by the transition of the flux measurements from the very outer parts of the group/IGL to the sky, until eventually the sky is the dominant source. Beyond this region the error bars become untrustworthy ($> \pm 0.2$~mag) so we could not extract any reliable conclusion. As a result, these outer colour values cannot be used for the IGL stellar population study. 

Overall, we see a gradient in the group colour profiles towards the blue in the transition between the core group region and the IGL. Then, the IGL shows a flat colour distribution.

By averaging the colour values of the profiles of the core galaxies of the group at SMA$<8$~arcsec (SMA$<30$~kpc), and proceeding in the same way for the IGL dominated region at $15<$SMA$<41$~arcsec ($50<$SMA$<140$~kpc), we measure the following colour values: the group $g-i =1.14 \pm 0.02$~mag, $g-r=0.78 \pm 0.02$~mag, and $r-i =0.35 \pm 0.03$~mag; while the IGL $g-i=0.92 \pm 0.01$~mag, $g-r=0.60 \pm 0.01$~mag, and $r-i=0.32 \pm 0.01$~mag. These two regions are indicated with the shaded grey areas in Figure~\ref{fig:IGLcolor}.

\subsection{Fraction of IGL} \label{subsec:IGL_fraction}

\begin{table*}
	\centering
	\caption{Fractions of IGL (top) and BCG+IGL (middle) over the total light of the group of galaxies at each band as indicated in the column names. The first three columns show the fractions for each of the two methods used in this work: 2D composite models (from Section~\ref{subsubsec:IGL_frac_2Dmod}) or surface brightness (SB) cuts (from Section~\ref{subsubsec:IGL_frac_SBcut}), and the last three columns are their associated luminosities. The bottom row shows the total luminosity of the 400138 group of galaxies in each band. (*) Surface brightness cuts values are in mag~arcsec$^{-2}$ units. }
	\label{tab:IGLfraction}
    \begin{tabular}{lcccccc}
        \hline
          Method (IGL frac.) & IGL$_{g}^{\rm{frac}}$ & IGL$_{r}^{\rm{frac}}$ & IGL$_{i}^{\rm{frac}}$ & $L_{g} ($L$_{\sun})$ & $L_{r} ($L$_{\sun})$ & $L_{i} ($L$_{\sun})$  \\
        \hline
            2D Comp. model & $36.5 \pm 2.2\%$ & $30.5 \pm 1.4\%$ & $29.8 \pm 1.1\%$ & $\sim 8.8 \times 10^{10}$  & $\sim 9.4 \times 10^{10}$  & $\sim 1.1 \times 10^{11}$ \\
            SB cut $\mu > 26^{(*)}$ & $34.0 \pm 2.3\%$ & $8.6 \pm 2.4\%$ & $4.0 \pm 2.8\%$ & $\sim 8.2 \times 10^{10}$  & $\sim 2.7 \times 10^{10}$  & $\sim 1.5 \times 10^{10}$ \\
            SB cut $\mu > 26.5^{(*)}$ & $14.6 \pm 3.2\%$ & $3.5 \pm 3.7\%$ & $1.6 \pm 4.3\%$ & $\sim 3.5 \times 10^{10}$  & $\sim 1.1 \times 10^{10}$  & $\sim 6.1 \times 10^{9}$\\
            &  &  &  &  &  & \\
        \hline
          Method (BCG+IGL frac.)  &  &  &  \\
        \hline
            2D Comp. model & $54.9 \pm 1.1\%$ & $51.5 \pm 0.6\%$ & $50.7 \pm 0.5\%$ & $\sim 1.3 \times 10^{11}$  & $\sim 1.6 \times 10^{11}$  & $\sim 1.9 \times 10^{11}$ \\
            &  &  &  &  &  & \\
        \hline
            400138 Group & -- & -- & -- & $\sim 2.4 \times 10^{11}$  & $\sim 3.1 \times 10^{11}$  & $\sim 3.8 \times 10^{11}$ \\
        \hline
    \end{tabular}
\end{table*}

The IGL fraction is a measurement of how much the IGL contributes to the total light of the group. This total light is considered as the sum of the flux coming from all the galaxy members in the group, including the BCG and any diffuse light around them. We calculated that total flux by summing the light detected within three apertures. The largest aperture is an ellipse centred on the centre of mass of the system composed by the galaxies in the core of the group (see details in Section~\ref{subsec:IGL_SBP}) with a SMA of 80~arcsec (275~kpc), an ellipticity of 0.52, and PA of 194~deg. This ellipse covers the light from the BCG, 1660730, 1660646, 1660615, and the diffuse component. Two more circular apertures are used centred on galaxy members 2301069 and 1660545, with radii of 6.7~arcsec (23~kpc) and 7.6~arcsec (26~kpc) respectively. We used the \textit{group mask} (see Section~\ref{subsec:mask}) over the star-subtracted images to allow for all the system's light. When measuring flux values throughout this work, we do not include flux below the surface brightness limit of our data at 1$\sigma$ within 10~$\times$~10~arcsec$^{2}$ boxes. By doing that, we minimise the background contamination. These surface brightness limits are as follows: $\mu _{g} ^{1\sigma \rm{lim}}$=31.95~mag~arcsec$^{-2}$; $\mu _{r} ^{1\sigma \rm{lim}}$=31.01~mag~arcsec$^{-2}$; $\mu _{i} ^{1\sigma \rm{lim}}$=30.61~mag~arcsec$^{-2}$

Our main challenge is to find a way to properly and systematically separate the light between different astronomical structures. Indeed, this is a well known problem that does not yet have a unique solution \citep{Montes2022}. When dealing with low surface brightness features such as the IGL, this issue is even harder to address as small inaccuracies in the flux separation imply large uncertainties in the IGL estimation due to its faint nature. This has led to a diverse range of ways of defining and extracting the IGL component. Some of those methods are more widely used than others so we have estimated the IGL fraction of 400138 using two of the most common ones. We choose this approach to contribute to uniformity in the field, and to facilitate further comparisons with earlier research.

\subsubsection{Method 1: Fraction of IGL after modelling and subtracting the group galaxies} \label{subsubsec:IGL_frac_2Dmod}

The first method measures the light of the IGL directly from the 2D star- and galaxy-subtracted images masked with the \textit{diffuse light mask}. This is, the images containing only the light from the diffuse component (see details in Section~\ref{subsubsec:2Dmodels}). The bottom panels of Figure~\ref{fig:GroupIm} show those images in each band. We obtained the amount of IGL using the large central ellipse aperture discussed in Section~\ref{subsec:IGL_fraction}. By dividing the IGL flux by the total flux of the group in each band, we obtained the following IGL fraction values (also indicated in Table~\ref{tab:IGLfraction} along with the corresponding luminosity values): IGL$_{g}^{\rm{frac}} = 36.5 \pm 2.2\%$, IGL$_{r}^{\rm{frac}} = 30.5 \pm 1.4\%$, IGL$_{i}^{\rm{frac}} = 29.8 \pm 1.1\%$.

We then measured the fraction of light of the brightest central galaxy (BCG, see Figure~\ref{fig:GroupIm}) and IGL, this is, the BCG+IGL fraction. As there are 3 galaxies in the core of 400138 with similar mass and sizes, we considered as the BCG the galaxy identified as such in GAMA \citep[][]{Robotham2011}. To obtain the BCG+IGL fraction, we followed the same procedure as above but over the star-subtracted images (i.e. with the core galaxies) masked with the \textit{BCG mask} (allowing for the BCG light only). We estimated the following BCG+IGL fractions in each band (see also Table~\ref{tab:IGLfraction}): BCG+IGL$_{g}^{\rm{frac}} = 54.9 \pm 1.1\%$, BCG+IGL$_{r}^{\rm{frac}} = 51.5 \pm 0.6\%$, BCG+IGL$_{i}^{\rm{frac}} = 50.7 \pm 0.5\%$.

\subsubsection{Method 2: Fraction of IGL from surface brightness cuts} \label{subsubsec:IGL_frac_SBcut}

The second method to obtain the IGL fraction is to simply apply a surface brightness cut assuming that all the flux fainter than that threshold belongs to the IGL component. That surface brightness cut is typically $\mu _{V}>$~26.5~mag~arcsec$^{-2}$ \citep[e.g.][]{Feldmeier2004, Rudick2011, Cui2014} and is extracted from the images with the core group galaxies PSF deconvolved (see details in Section~\ref{subsubsec:2Dmodels}). The IGL flux in each band is then obtained by adding the flux from all the pixels below the surface brightness cut and above the data threshold established (the surface brightness limit of our data at 1$\sigma$ within 10~$\times$~10~arcsec$^{2}$ boxes). For a more comprehensive analysis, we applied the two most extended surface brightness cuts, 26 and 26.5~mag~arcsec$^{-2}$. Thus, the resulting IGL fractions for a cut at 26~mag~arcsec$^{-2}$ are $\sim 34\%$, $\sim 9\%$, and $\sim 4\%$ ($\pm 3\%$) in the $g$, $r$, and $i$ band respectively. And for those when cutting at 26.5~mag~arcsec$^{-2}$ are $\sim 15\%$, $\sim 4\%$, and $\sim 2\%$ ($\pm 4\%$) in the $g$, $r$, and $i$ band respectively. These results are summarised in Table~\ref{tab:IGLfraction}. 

As the group is located at $z\sim0.21$ the rest-frame set of surface brightness cuts are converted to the corresponding observed cuts by applying the redshift surface brightness dimming correction and the K-correction \citep[from the GAMA catalogue;][]{Loveday2012} in each band. It is important to take into account that this method considers the IGL as the diffuse light associated with all of the galaxies in the group, rather than just that around the BCG as in the method applied in Section~\ref{subsubsec:IGL_frac_2Dmod}. Regarding the PSF influence, there is no significant difference in the IGL fraction estimation: when accounting for the PSF effect the IGL fraction differs $\sim 1 \%$ in the most affected case (this is, for a surface brightness cut at 26~mag~arcsec$^{-2}$ in $g$ band). This value is within the errors of the IGL estimation.

\section{Discussion} \label{sec:discussion}

In this pilot work we have used HSC-PDR2 data to detect IGL in the GAMA group of galaxies 400138 ($M_{\rm{dyn}}= 1.34 \times 10^{13} $M$_{\sun}$) at $z\sim0.21$ for the first time. This system is a loose group although the core region is a dense environment composed of three galaxy members. We have measured the IGL to an extent of SMA$\sim 275$~kpc in the $g$, $r$, and $i$ bands. We also studied the colour distribution of the system to SMA$\sim 170$~kpc after correcting for the PSF effects. The IGL shows a flat radial colour distribution with mean values of $g-i=0.92$, $g-r=0.60$, and $r-i=0.32$ ($\pm 0.01$~mag). We find values for the fraction of IGL flux relative to the group from $\sim 2 \%$ to $\sim 36 \%$ depending on the method and wavelength of measurement. In this Section we discuss these results and compare them with previous observational works and theoretical predictions to infer the dominant physical process in the formation of the IGL component.

\subsection{IGL extent and radial distribution} \label{subsec:IGL_extent}

We have extracted surface brightness profiles of the IGL in the GAMA group 400138 with reliable detection down to the surface brightness limits of the $g$, $r$, and $i$ bands (see Figure~\ref{fig:IGLsbp}). Those profiles measured the IGL component to a radial distance of SMA$\sim 275$~kpc ($\sim 80$~arcsec) from the centre of the group for the aforementioned bands. From those profiles, we extracted the ($g-i$), ($g-r$), and ($r-i$) colour profiles to an extent of SMA$\sim 170$~kpc ($\sim 50$~arcsec). In comparison with previous works, our IGL detection is slightly larger than that of \citet[][]{Ragusa2021} in the Hickson Compact Group of galaxies \citep[HCG;][]{Hickson1997} HCG 86, with an IGL extension of $\sim 170$~kpc in $r$ band from the centre of HCG86A and a ($g-r$) colour profile out to 160~kpc. Prior to the study of \citet[][]{Ragusa2021} IGL detections extended to radii between 30 and 100~kpc from the centre of the group \citep[e.g.][]{DaRocha2005, DaRocha2008, DeMaio2018, Poliakov2021}. Our IGL measurements are the most extended to date. This is remarkable considering that our system is relatively low mass ($M_{\rm{dyn}}=1.34 \pm 0.5 \times 10^{13} $M$_{\sun}$) and is located at $z\sim 0.21$. In addition, 400138 is not a compact group.

As described in Section~\ref{subsec:IGL_image}, the IGL radial distribution of the group 400138 is asymmetric. It extends up to the location of the galaxy group member 1660615. In this asymmetric region, the IGL shows a component towards the north that spreads out well beyond the core of the group. This may suggest an interaction event between a group member, likely 1660615, and the galaxies in the core region of the group. According to N-body simulations of the formation and evolution of substructures within the ICL \citep[referred to as ``streams'';][]{Rudick2009}, $\sim40\%$ of the ICL is generated through massive, dynamically cold streams. \citet{Rudick2009} found that the fraction of ICL that is produced in the form of streams is greater when galaxies are interacting in the group environment than in the cluster stage. HSC observations suggest that the 400138 galaxy group is still in an active (non relaxed) phase in the lifetime of the system. When such ICL (or IGL) substructures are found in the core of the system they only live for up to $\sim 1$~Gyr \citep{Rudick2009}. The reason for such a quick decay is that their dynamical times are short because of the tidal field of the cluster. However, the core of our group is not as dense as a cluster core. Therefore, the stream may not be as easily disturbed as in a cluster and it might survive for a longer period of time. We can roughly estimate the dynamical time of the IGL stream in 400138 from Eq.~3 in \citet{Rudick2009}. To do that, we assumed that the stream is at time $t_{\rm{max}}$ at which it has reached its maximum radial distance from the centre of the group (i.e. mean position of the stream at time $t_{\rm{max}}$ of $r \sim 200$~kpc) and that the enclosed mass is $M_{\rm{encl}} \sim 1.27 \times 10^{13} M_{\odot}$. This is, the group dynamical mass minus the individual halo masses of the members 1660545 and 2301069. The individual halo masses were obtained from the empirical relation between the mass of the host dark matter halo and the total stellar mass within the halo from \citet{Huang2020}, using the stellar masses in \citet[][see Section~\ref{subsec:GAMA}]{Taylor2011}. We obtained a dynamical time of 0.6~Gyr, which means a decay time for the IGL stream of at least 0.9~Gyr (1.5 times the dynamical time), in agreement with the time scales predicted by \citet{Rudick2009} for clusters. In summary, 400138 seems to have experienced an interaction event at least $\sim 0.9$~Gyr ago between the core group galaxies and a galaxy member, probably 1660615, producing the stream-like substructure observed in the IGL component.



\begin{figure*}
\begin{center}
\includegraphics[width=14cm]{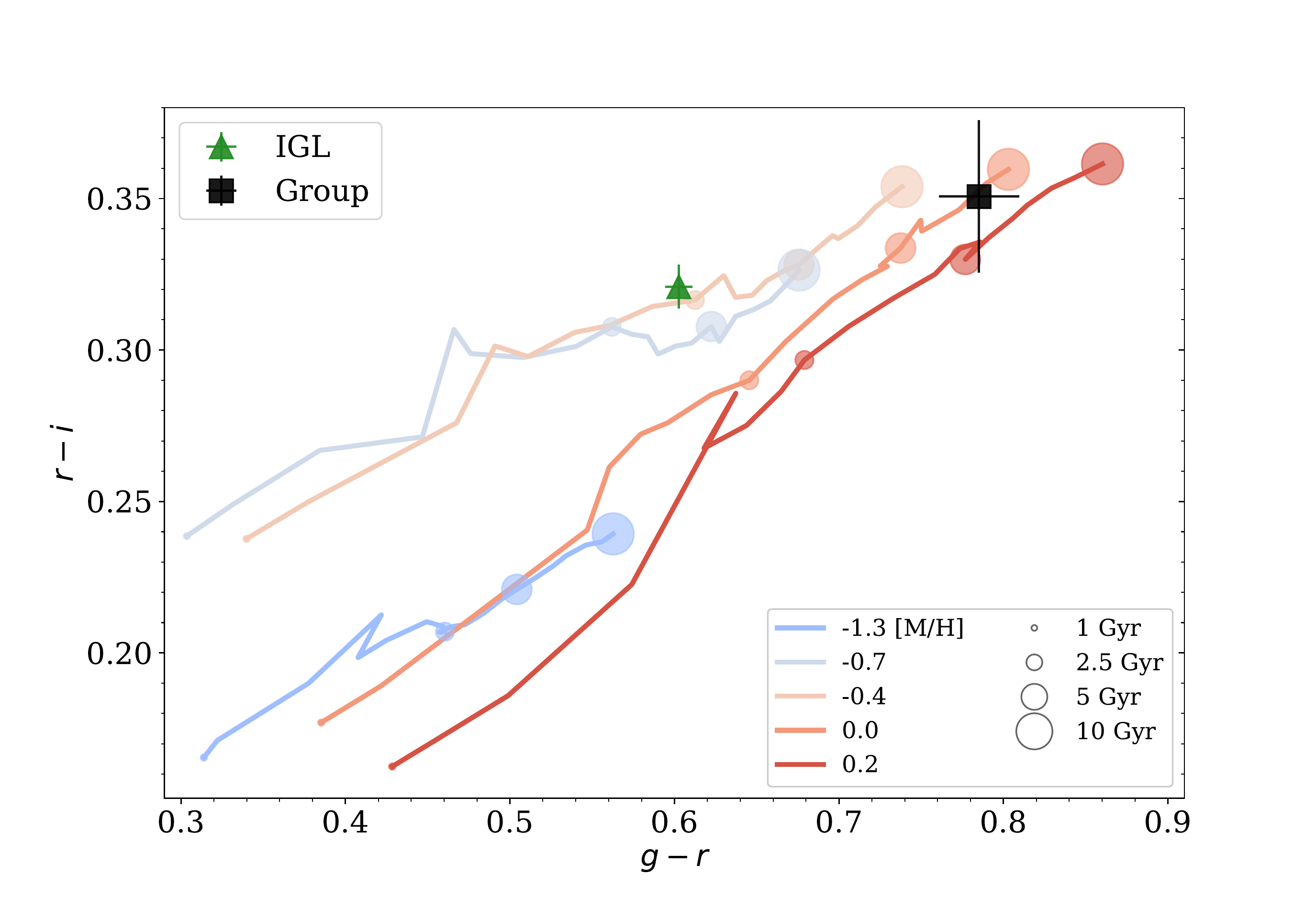}
\caption{$(g-r)-(r-i)$ colour diagram of the 3 core group galaxies (Group; black square) and the IGL component (IGL; green triangle). Overplotted are the age (circles) and metallicity tracks (solid lines) from the E-MILES stellar population models \citep{Vazdekis2016}. We find the group galaxies to be consistent with 7--10~Gyr age and solar metallicity ([Fe/H]~$ \sim 0$), while the IGL stellar populations are younger (2--2.5~Gyr) and less metal-rich ([Fe/H]~$ \sim -$0.4). 
  \label{fig:color-color}}
\end{center}
\end{figure*}

\subsection{Stellar populations in the IGL component} \label{subsec:IGL_StelPop}

Colour studies of groups of galaxies and their IGL provide information on the properties of the stellar populations that they host, which are related to the mass assembly history of their host system \citep[e.g,][]{Montes2014, Morishita2017, DeMaio2018, Montes2018, Contini2019a, Montes2021a}. Figure~\ref{fig:IGLcolor} shows a colour gradient in the group profiles towards the blue in the transition region between the core group region and the IGL (SMA$<50$~kpc). Then, the IGL shows an almost flat colour distribution up to SMA$\sim 140$~kpc. These features have previously been found in clusters and a group of galaxies \citep[e.g.,][]{Montes2014, Iodice2020} while others have observed colour gradients in the opposite sense \citep[e.g.,][]{Spavone2018, Ragusa2021}. According to the proposed formation mechanisms for the diffuse component within dense systems of galaxies \citep[see review by][]{Montes2019a}, a colour gradient towards bluer wavelengths is indicative of younger and less metal-rich stellar populations in the IGL than that of the core group galaxies. Semi-analytic models of galaxy formation in clusters and groups suggest this can be produced by partial tidal stripping of satellite galaxies less massive than those in the core \citep[][]{Contini2014}. The flat colours in the IGL dominant region imply the presence of accreted material or ongoing mergers at those distances that mix the stellar populations in the region and smooth out any previous gradient. This is consistent with our observation of a stream in the IGL as discussed in Section~\ref{subsec:IGL_extent}.


From the colour profiles in Figure~\ref{fig:IGLcolor}, we extracted the average colours of the core galaxies of the group at SMA$<30$~kpc, and those of the IGL in the region at $50<$SMA$<140$~kpc. These mean IGL values (Section~\ref{subsec:IGL_color}) are in agreement with the semi-analytic models of \citet{Contini2019a} for the IGL in systems with halo masses of $10^{13}-10^{15} $M$_{\sun} h^{-1}$ at redshift $0<z<0.5$. 

Figure~\ref{fig:color-color} shows the mean colours for the IGL and the core group members, in a $(g-r)-(r-i)$ colour diagram together with the age and metallicity\footnote{Please note that the relation between the total metallicity and iron metallicity is [Fe/H]=[M/H]-0.75[Mg/Fe] but we assume that [Fe/H] = [M/H] as we use ``base'' stellar population models. These models use scaled-solar isochrones but do not consider the specific abundance element ratio of the stellar spectra. Base models follow the abundance pattern of the Milky Way as a function of metallicity and are scaled-solar around solar metallicity \citep{Vazdekis2016}.} tracks derived from the E-MILES stellar population models \citep{Vazdekis2016} assuming a \citet{Chabrier2003} initial mass function (IMF) and Padova+00 isochrones \citep{Girardi2000}. The IGL stellar populations are younger (2--2.5~Gyr) and less metal-rich ([Fe/H]~$ \sim -$0.4) than those of the most massive galaxy members (7--10~Gyr, [Fe/H]~$ \sim$~[Fe/H]$_{\sun}$), in agreement with predictions by \citet{Contini2019a}. This supports the formation of the IGL from tidal stripping of infalling satellite (less massive) galaxies with younger and less metal-rich stellar populations than the group members. Considering the difference between the most massive (core) galaxies in the group and the IGL (4.5--8~Gyr), and the estimated age values for each component, we can estimate the formation time of their stars. The stellar populations in the core group are compatible with a formation time at $z \sim 2-3$ (i.e., $\sim 10-11$~Gyr ago), and, in consequence, the bulk of the faint component would have been formed relatively recently, at $z \sim 0.8-1.2$ (i.e., $\sim 6.5-8.5$~Gyr ago). 

\citealt{Montes2014} were the first to find such a colour gradient in the ICL of the massive Abell Cluster 2744 at $z=0.3$. Their ICL was $6 \pm 3$~Gyr younger than the average age of the most massive galaxies of the cluster and had solar metallicity ([Fe/H]~$ \sim 0.018$). \citet{Montes2018} measured the ICL in six massive intermediate redshift ($0.3 < z < 0.6$) clusters from the Hubble Frontier Fields (HFF). They obtained an average ICL metallicity of [Fe/H]$_{\rm {ICL}} \sim -$0.5 and mean stellar ages of the ICL between 2 and 6~Gyr younger than the most massive galaxies of the clusters \citep[see also][]{Morishita2017}. These results are in good agreement with ours but each of these studies were done in massive clusters with masses $>10^{15} M_{\odot}$. The observed stellar populations of the IGL are also equivalent to the integral field spectroscopic results from \citet{Edwards2020}, where the ICL within 23 massive clusters at $0.03<z<0.07$ was best modelled with younger and less metal-rich stars than those in the BCGs. Age differences of 2--6~Gyr between the ICL and the BCG have been found in intermediate-redshift clusters \citep{Toledo2011, Adami2016}, in good agreement with our results for 400138.

The metallicity values of the faint component in 400138 are consistent with the mean value of the outskirts of the Milky Way ([Fe/H]$_{\rm {ICL}} \sim $[Fe/H]$_{\rm {MW out.}} \sim -$0.5; \citealt{Cheng2012a}), and similar to those reported by \citet{Montes2018}. According to the stellar mass-metallicity relation \citep[e.g.][]{Gallazzi2009}, our IGL metallicity corresponds to the metallicities of progenitor galaxies with $M_{\rm{prog}} \sim 0.8 \times 10^{10} M_{\odot}$. Considering that the stars in the outer regions of galaxies are more easily stripped, and taking into account the intrinsic metallicity gradients within these systems, galaxies more massive than $\sim 10^{10} M_{\odot}$ are still good candidates to be the primordial source of stars for the IGL observed here. That could be the group galaxy 1660615 ($M_{*} \sim 2.4 \times 10^{10} M_{\odot}$; \citealt[][]{Taylor2011}) or even since dissolved Milky Way-like galaxies ($6.43 \pm 0.63 \times 10^{10}  M_{\odot}$; \citealt[][]{McMillan2011}). The colours in the outer regions of 1660615 have a mean metallicity [Fe/H]$_{1660615} \sim -$0.4, similar to that of the outer part of the IGL. In the case of the core galaxies, the colours in their outer parts and the derived ages and metallicities, match with that of the inner part of the IGL within the uncertainties of the method.




\subsection{Fraction of IGL in the group} \label{subsec:IGL_fraction_Discus}

In Section~\ref{subsec:IGL_fraction} we found that the fraction of IGL flux relative to the group that ranges between $\sim$2$\%$ and $\sim$36$\%$. The wide range of IGL fraction values is the result of the different methodologies applied, which rely on the different approaches to what is defined as the diffuse component of a group or cluster of galaxies. N-body simulations of clusters of galaxies ($0.8-6.5 \times 10^{14} $M$_{\sun}$) by \citet{Rudick2011} predicted that ICL fractions at $z \sim 0.2$ range from $5\%$ to $20\%$ using a variety of measurement techniques, including surface brightness cuts, finding that different methods can change the measured ICL fraction by up to a factor of two within a single cluster. It is noteworthy that the IGL fractions obtained from the 2D composite models are systematically larger than those from the surface brightness cuts (see Table~\ref{tab:IGLfraction}).

\citet{SommerLarsen2006}'s cosmological simulations of galaxy groups predicted that between 12$\%$ and 45$\%$ of the light in these systems belongs to the IGL by defining the IGL stars as those located between the BCG outer radius and the group virial radius but not inside of the tidal radius of any galaxy in the group. Cosmological hydrodynamical simulations by \citet{Murante2004, Murante2007} of galaxy clusters ($\sim 10^{14} $M$_{\sun} - 10^{15} $M$_{\sun} h^{-1}$) estimated a fraction of stars unbounded from other galaxies in the systems of 20--50$\%$ and $\sim$15--30$\%$ respectively. However, \citet[][]{Puchwein2010} found a consistent ICL fraction of $\sim 45 \%$ for halo masses between $\sim 10^{13} $M$_{\sun} h^{-1}$ and $\sim 10^{15} $M$_{\sun} h^{-1}$ using four different methods to identify the ICL in high-resolution hydrodynamical simulations. \citet{Contini2014} found ICL fractions between 10$\%$ and 40$\%$, independent of the halo mass. They defined the ICL differently for each of the three semi-analytic models of galaxy formation in clusters and groups of galaxies they applied. \citet[][]{Tang2018} applied the surface brightness cut method at $\mu _{V}>$~26.5~mag~arcsec$^{-2}$ over mock images from cosmological hydrodynamical simulation to extract ICL fractions. They measured values of $\sim$16--20$\%$ for systems at redshift $z \sim 0.2$ and a halo mass of $\sim 10^{13}~$M$_{\sun}$. Recently, IGL and ICL fractions obtained using the \textsc{velociraptor} galaxy finding algorithm \citep{Canas2019} from the Horizon-AGN cosmological hydrodynamical simulation \citep{Dubois2014}, derived a narrower fraction range of the IGL component of $\sim$10--20$\%$ \citep{Canas2020}. In summary, there is not agreement in the IGL and ICL fraction predictions due to the different approaches used to measure and define these components, as well as the wide variety of halo masses and evolutionary states of the systems studied. The development of a common definition is fundamental for further analysis in the field.

Assuming an average IGL fraction of $~30 \%$, our group of galaxies is compatible with the predictions by \citet{Contini2014} for an IGL formation model \citep[derived from][]{Villalobos2012} that includes continuous stripping of satellite galaxies within a similar halo mass group $\sim 10^{13}~$M$_{\sun}$. These simulations estimate that in a group of galaxies with mass similar to 400138 and its BCG, the continuous stripping of massive satellite galaxies, produces an IGL fraction of $~30 \%$ with metallicity of -0.5~$ < $[Fe/H]$ < -$0.32 (also compatible with our results as reported in Section~\ref{subsec:IGL_StelPop}).


One of the very few studies of the IGL properties in a loose group of galaxies is the analysis by \citet{Spavone2018} of the group NGC~5018. This system is less massive than 400138, with a virial mass of $\sim 4 \times 10^{12} $M$_{\sun}$, but presents a large fraction of IGL of $\sim 41\%$ ($g$ band), obtained from multi-component fits to the azimuthally averaged surface brightness profiles. The IGL fraction from \citet{Spavone2018} is compatible within the uncertainties with our IGL fraction of $\sim 36\%$ in $g$ band obtained with the 2D models (and that from the surface brightness cut at $\mu _{g}>$~26~mag~arcsec$^{-2}$, $\sim 34\%$). Both of these loose groups have high IGL fractions and signs of interactions including tidal stellar features although with only two systems the sample is too limited to derive any conclusions. In the case of 400138, we find an extended faint tail, while in NGC~5018 there are galaxies with disrupted morphologies, especially in their outskirts, and tidal stellar tails are also observed \citep{Spavone2018}. 


The IGL stellar mass of 400138 is $M^{\rm{IGL}}_{i} \sim 1.1 \pm 0.5 \times 10^{11} M_{\odot}$. We derived this mass following the empirical relation between $(g - i)$ colour and mass-to-light ratio $M_{*}/L_{i}$ from \citet{Taylor2011} assuming a \citet{Chabrier2003} IMF. We considered an IGL fraction of $~30 \%$ obtained from the 2D composite model method in $i$ band (Section~\ref{subsubsec:IGL_frac_2Dmod}), and the mean IGL colour $g-i=0.92 \pm 0.01$~mag. Assuming a Milky Way mass of $6.43 \pm 0.63 \times 10^{10}  M_{\odot}$, it would require the total disruption of 1--2 galaxies similar to the Milky Way to produce the IGL we are detecting in our group. If we consider that galaxies similar to group member 1660615 ($M_{*} \sim 2 \times 10^{10}  M_{\odot}$) are the potential progenitors, the IGL formation in 400138 would need the total disruption of 4--5 of these galaxies.



\begin{figure}
\begin{center}
\includegraphics[width=\columnwidth]{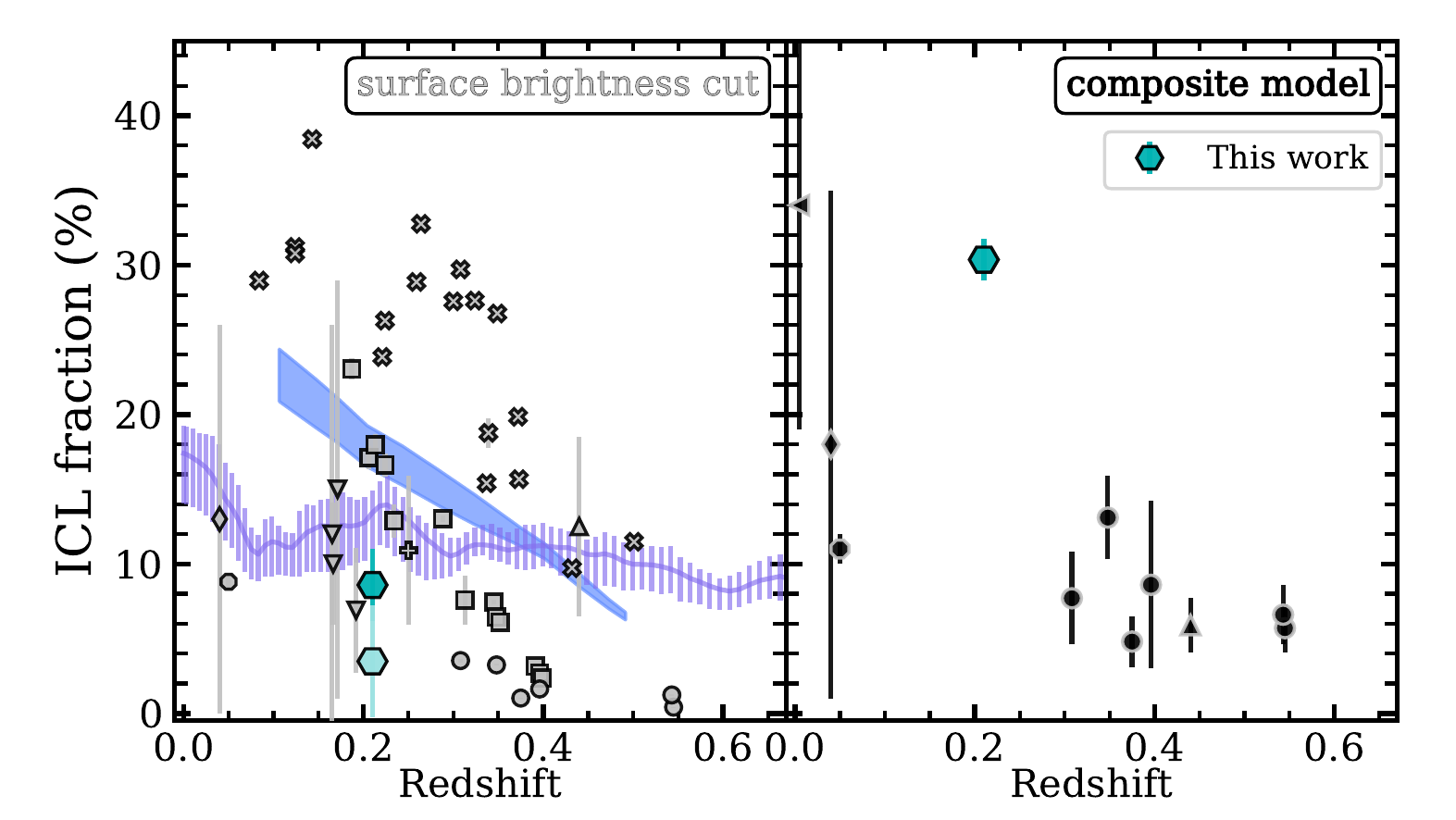}
\caption{IGL fraction in this work (teal hexagons, from $r$-band values) as a function of redshift compared with IGL and ICL fractions from the literature (black and grey points; updated from \citealt{Montes2022}). The \textit{left} panel shows the fractions derived using a surface brightness cut, where the purple line and errors correspond to the simulations of \citet{Rudick2011} for $\mu _{V}>$~26~mag~arcsec$^{-2}$, and the blue shaded region illustrates the predictions from \citet{Tang2018} for $\mu _{V}>$~26.5~mag~arcsec$^{-2}$. The \textit{right} panel shows IGL and ICL fractions obtained from composite models to fit the BCG and ICL, separately. The different markers, and their corresponding errors, indicate the studies
from which the fractions were taken from. 'x': \citet{Furnell2021}, squares: \citet{Burke2015}, circles: \citet{Montes2018}, diamond: \citet{Kluge2021}, octagon: \citet{Montes2021a}, cross: \citet{Zibetti2005}, star: \citet{Ko2018}, triangle: \citet{Presotto2014}, down-pointing triangles: \citet{Feldmeier2004}, left-pointing triangle: \citet{Spavone2020}.
  \label{fig:IGLfraction}}
\end{center}
\end{figure}

In Figure~\ref{fig:IGLfraction}, we show our IGL fraction measurements in comparison with previous ICL/IGL studies as a function of redshift separated in two panels depending on the methodology applied. We do not see a clear correlation between the ICL fraction in groups and clusters of galaxies and redshift. Our result obtained from the 2D composite model method (details in Section~\ref{subsubsec:IGL_frac_2Dmod}) suggests an increase of the faint component at lower $z$. This is in agreement with the Fornax Cluster analysis by \citet{Spavone2020}, although with large uncertainties, but contrary to the general trend derived from massive clusters \citep{Montes2018, Kluge2021, Montes2021a}. However, these IGL and ICL fractions have also been obtained from different photometric bands so potentially trace different stellar populations \citep[see the more detailed discussion on this topic in][]{Montes2022}. Using the surface brightness cut method, some previous studies have pointed to a possible time evolution in the ICL, remaining roughly constant before $z\sim 0.6$, and then the fraction of light relative to that of the total group light increasing rapidly from $z \lesssim 0.6$ to present time \citep[e.g.][]{Tang2018}. Our measurements are not consistent with that increase at lower redshifts and are consistent with lower ICL fractions such as those obtained by \citet[][]{Feldmeier2004} and \citet[][]{Montes2018} in clusters of galaxies. More data points are clearly needed, particularly for groups of galaxies and for systems at higher redshifts ($z > 0.6$), in order to build a complete picture of the IGL evolution.

The IGL fractions for 400138 (Table~\ref{tab:IGLfraction}) show that they decrease towards redder wavelengths. That is an indication of a younger stellar population in the IGL than in the most massive galaxies in the group. This was also found by \citet{JimenezTeja2018} and \citet{JimenezTeja2019} who measured higher ICL fractions at blue wavelengths in merging clusters (dynamically active) compared to non-merging clusters. This is consistent with our stellar IGL population analysis (Section~\ref{subsec:IGL_StelPop}) where we argue that the excess of lower-metallicity/younger stars have been stripped from the outskirts of the infalling galaxies during a merger event. This adds further evidence that 400138 is an active system with ongoing merging events. 

\begin{figure}
\begin{center}
\includegraphics[width=\columnwidth]{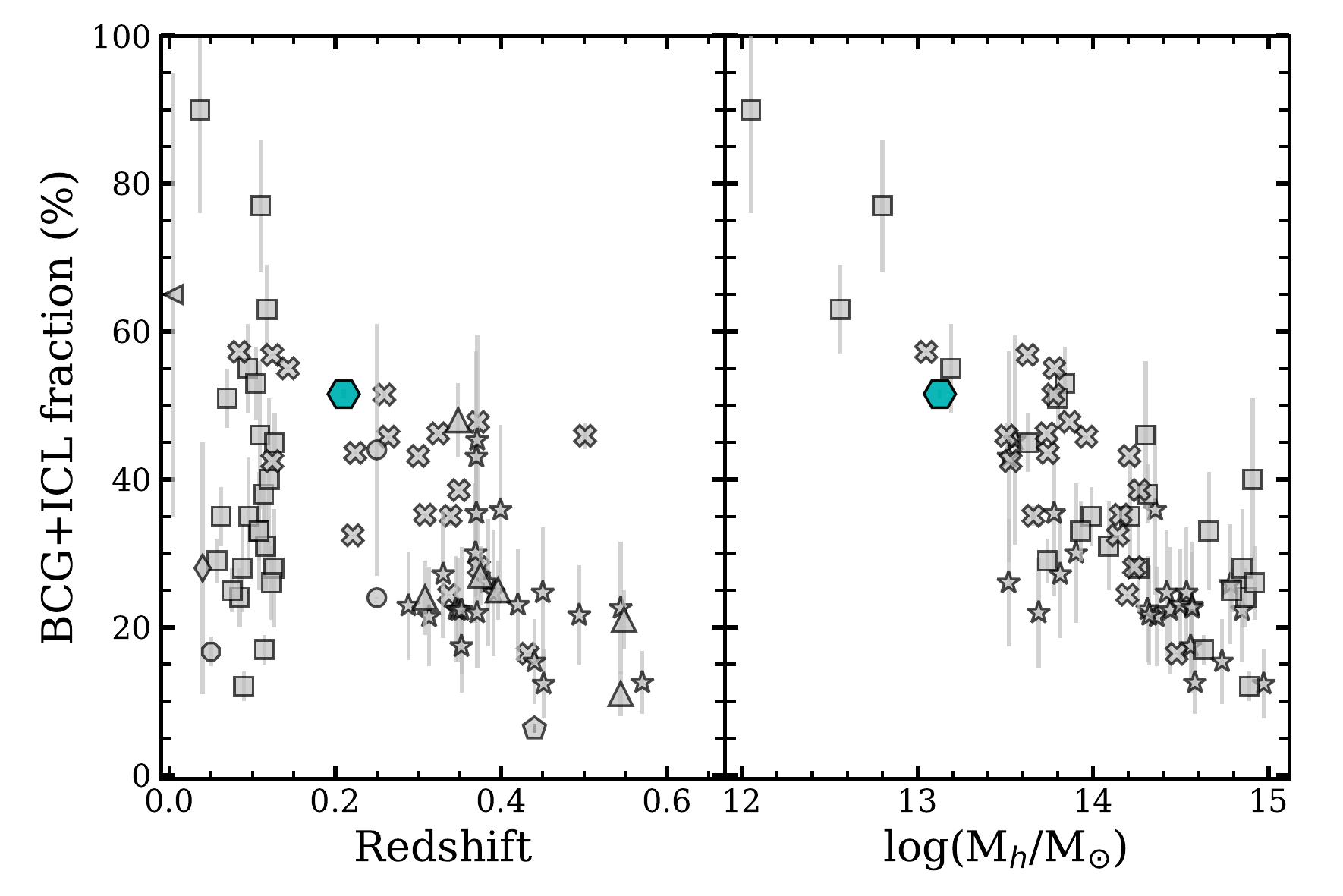}
\caption[]{BCG+ICL fraction as a function of redshift (\textit{left}) and as a function of the total mass of the system (\textit{right}) of our group (cyan hexagon) and previous works (grey points; updated from \citealt{Montes2022}). The different markers, and their corresponding errors, indicate the studies from which the fractions were taken from. Squares: \citet{Gonzalez2007}, stars: \citet{DeMaio2018}\footnotemark, 'x': \citet{Furnell2021},  triangles: \citet{Morishita2017}, diamond: \citet{Kluge2021}, pentagon: \citet{Presotto2014}, octagon: \citet{Montes2021a}, left-pointing triangle: \citet{Spavone2020}, circles: \citet{Zhang2019} and \citet{SampaioSantos2021}. \label{fig:BCGIGLfraction}}
\end{center}
\end{figure}

The combined BCG+IGL flux fraction relative to that of the total group in $r$ band is $\sim$51$\%$. In Figure~\ref{fig:BCGIGLfraction} we compare this result with previous research as a function of the redshift ($z\sim0.21$), and total mass ($M_{\rm{dyn}}= 1.34 \pm 0.5 \times 10^{13} $M$_{\sun}$) of the system. We find good agreement with previous works. Firstly, our BCG+IGL fraction results concur with those from the study of clusters of galaxies at similar redshifts \citep[e.g.][]{Furnell2021}. We also find similar BCG+IGL fraction measurements to observed nearby groups of galaxies with similar masses \citep[e.g.][]{Gonzalez2007}. However, we note that these are slightly smaller values than those predicted by \citet[][60--90$\%$]{Contini2014} but within the range estimated by \citet[][50--60$\%$]{Puchwein2010} using high-resolution hydrodynamical simulations. On the other hand, as previously reported by \citet{Montes2022}, the strong negative correlation between the BCG+ICL fraction and the halo mass indicates that when the system halo mass increases, there is more mass held in the satellite galaxies \citep[e.g.][]{Brough2008}, decreasing the BCG+ICL fraction. Without more data it is not possible to disentangle any conclusions from the BCG+IGL fraction relation with system mass or redshift. 

\footnotetext{We derived the BCG+IGL fractions from \citet{DeMaio2018} because they do not provide them. In \citet{DeMaio2018}, they gave the stellar mass of BCG+ICL inside 100~kpc from the centre of the BCG and the $M_{500}$ of the clusters. To derive the stellar mass of the clusters, we used the relationship between the total stellar mass of a cluster and its $M_{500}$ given in \citet{Kravtsov2018}. Therefore, the fractions plotted in Figure~\ref{fig:BCGIGLfraction} are actually the BCG+ICL within 100~kpc divided by the total stellar mass of the system. We might be missing a significant fraction of the BCG+ICL \citep{Gonzalez2021} and therefore, we warn the reader that these fractions should be considered as a lower limit estimation.}

\subsection{Relation of the diffuse light with the ETGs-to-LTGs ratio}

\citet{DaRocha2008} used the ratio of early-type galaxies to late-type galaxies (ETG/LTG) as an indicator of the dynamical evolution of the group. They found a linear relation between the IGL fraction and the ETG/LTG fraction in their sample of compact groups \citep{DaRocha2005, DaRocha2008}. Recently, \citet{Ragusa2021} found the IGL fraction of the compact group HCG 86 to be consistent with this analysis.


According to the colour and morphology of each of our member galaxies and after visual inspection, we estimate an ETG-to-LTG ratio of 5. The 400138 group is dominated by early-type galaxies (five early-type and one late-type). Considering the IGL fraction obtained with the same method to those of \citet{DaRocha2008} and \citet{Ragusa2021}, i.e. $\sim$36$\%$ in $g$ band from the 2D composite model method, our value is consistent with the relation of \citet{DaRocha2008}, and falls within 2$\sigma$ of the relation in \citet{Ragusa2021}. This adds further evidence to their conclusion that larger amounts of IGL are found in more evolved groups, dominated by ETGs.

Furthermore, according to \citet[][]{Crossett2022}, the X-ray luminosity of our group of galaxies falls into their ``X-ray overluminous'' category, associated with a decrease in blue galaxies and a reduction in total group star formation. These X-ray overluminous groups are the more evolved systems and thus, in agreement with detection of a significant IGL fraction in our group.

\subsection{Further analysis of a larger sample of groups of galaxies} \label{subsec:FurtherAn_Discus}

The 10-year Legacy Survey of Space and Time (LSST), which will take place at the Vera C. Rubin Observatory \citep[][]{Ivezic2019} is expected to observe around one million galaxy groups up to redshift $z=1$. This will enable a statistically significant study of the properties of the ICL and IGL across halo mass and redshift \citep[e.g.][]{Montes2019a, Brough2020}. The HSC-PDR2 data are processed using a modified version of the LSST pipeline which, combined with the extraordinary depth, make HSC-PDR2 data the perfect starting point to prepare for the arrival of LSST data. The Python-based analysis methods used here have been developed with LSST in mind. Therefore, this pilot work prepares us for a more comprehensive analysis of the IGL properties in groups of galaxies first in HSC and later in the upcoming LSST data.

\section{Conclusions} \label{sec:conclusions}


This paper presents a pilot study to assess the potential of Hyper Suprime-Cam Public Data Release 2 (HSC-PDR2) images for the analysis of extended faint structures within groups of galaxies. We report the first IGL detection in the loose GAMA galaxy group 400138 ($M_{dyn}= 1.34 \times 10^{13} $M$_{\sun}$) located at $z\sim 0.21$. We have measured the IGL through surface brightness profiles in the $g$, $r$, and $i$ bands down to $\mu _{g} ^{\rm{lim}}$=30.76~mag~arcsec$^{-2}$ (3$\sigma$ within 10~$\times$~10~arcsec$^{2}$ boxes). The main conclusions from the analysis of the extent, radial distribution, colour, and fraction of the IGL of 400138, as well as the lessons learnt from this pilot study, are the following:

\begin{itemize}
    \item We measured the IGL flux to a radial distance of SMA$\sim$275~kpc and extracted the ($g-i$), ($g-r$), and ($r-i$) colour profiles to an extent of SMA$\sim$170~kpc from the centre of the group. This is the most extended IGL detection to date.  
    
    \item The IGL shows an asymmetric radial distribution with a stream substructure that extends from the core of the group towards the location of the group member 1660615 ([Fe/H]$_{1660615} \sim -$0.4). This stream covers an area of $\sim 314 \times 75$~kpc$^{2}$, and has a similar metallicity to 1660615. We estimate the interaction time between the core group galaxies and a group member, likely 1660615, to have been at least $\sim 0.9$~Gyr ago.
    
    \item The IGL shows a flat colour radial distribution at $50<$SMA$<140$~kpc with mean values of $g-i$=0.92, $g-r$=0.60, and $r-i$=0.32 ($\pm$0.01~mag). At SMA$<50$~kpc there is a negative colour gradient which means that the IGL stellar populations are younger (2--2.5~Gyr) and less metal-rich ([Fe/H]~$ \sim -$0.4) than the host group (7--10~Gyr, [Fe/H]~$\sim $[Fe/H]$_{\sun}$). 
    
    \item The fraction of IGL relative to the group light ranges from $\sim$2$\%$ to $\sim$36$\%$ depending on the method of measurement and wavelength (Table~\ref{tab:IGLfraction}). We do not find any correlation between the IGL fraction and redshift with the available data (Figure~\ref{fig:IGLfraction}). The IGL has a stellar mass of $M^{\rm{IGL}}_{i} \sim 1.1 \pm 0.5 \times 10^{11} M_{\odot}$. 
    
    
    
    \item The IGL fraction of 400138 is larger the bluer the observation wavelength, in agreement with \citet{JimenezTeja2018} and \citet{JimenezTeja2019} for the ICL in dynamically active clusters of galaxies. 
    
    \item Our analysis of the IGL fraction and stellar populations of 400138 suggests that this is an active group with ongoing merging events where the bulk of the IGL may have mainly been formed at $z \sim 0.8-1.2$ by stars stripped from 4--5 recently quenched satellite galaxies with similar mass and stellar population properties to 1660615 or 1--2 galaxies similar to the Milky Way during merging processes. 

    \item The estimated BCG+IGL fraction ($\sim$51$\%$ in $r$ band) is in agreement with previous analyses (Figure~\ref{fig:BCGIGLfraction}), including studies of more massive systems at intermediate redshift \citep{Furnell2021} and local groups within a similar halo mass range \citep{Gonzalez2007}. 
    
    \item We find that the early-type to late-type galaxy ratio suggests that 400138 is a more evolved system, dominated by ETGs, and its large amount of IGL is consistent with previous analysis of this relation. 
    
    \item This pilot study has allowed the detection and comprehensive analysis of the IGL in a GAMA group of galaxies using Ultradeep HSC-PDR2 data. Our method has been successfully tested. A follow up work will be performed in a sample of galaxy groups to study the properties of their IGL across mass ranges and redshifts. This will confirm the potential of this method to detect IGL in groups with lower IGL fractions as 400138 is potentially at the high end of the IGL distribution for its mass.
    
\end{itemize}

\section*{Acknowledgements}

We acknowledge constructive remarks by an anonymous referee that help to improve this manuscript. We thank Felipe Jim{\'e}nez-Ibarra and Ignacio Trujillo for useful discussions. This research was supported by the Australian Research Council Discovery Project DP190101943. Parts of this research were supported by the Australian Research Council Centre of Excellence for All Sky Astrophysics in 3 Dimensions (ASTRO 3D), through project number CE170100013. CML, RBG, MA, and RIS acknowledge support from the State Research Agency (AEI-MCINN) of the Spanish Ministry of Science and Innovation under the grant ``The structure and evolution of galaxies and their central region'' with reference PID2019-105602GB-I00/10.13039/501100011033. MA acknowledges the financial support from the Spanish Ministry of Science and Innovation and the European Union - NextGenerationEU through the Recovery and Resilience Facility project ICTS-MRR-2021-03-CEFCA. RIS acknowledges the funding by the Governments of Spain and Arag{\'o}n through the Fondo de Inversiones de Teruel; and the Spanish Ministry of Science, Innovation and Universities (MCIU/AEI/FEDER, UE) with grant PGC2018-097585-B-C21.

This paper is based on data collected at the Subaru Telescope and retrieved from the Hyper Suprime-Cam (HSC) data archive system, which is operated by the Subaru Telescope and Astronomy Data Center (ADC) at National Astronomical Observatory of Japan. Search and download data from Hyper HSC-SSP2 was done thanks to \texttt{unagi} Python module developed by Song Huang and contributors. We have used catalogues from GAMA, a joint European-Australasian project based around a spectroscopic campaign using the Anglo-Australian Telescope. The GAMA input catalogue is based on data taken from the Sloan Digital Sky Survey and the UKIRT Infrared Deep Sky Survey. Complementary imaging of the GAMA regions is being obtained by a number of independent survey programmes including GALEX MIS, VST KiDS, VISTA VIKING, WISE, Herschel-ATLAS, GMRT and ASKAP providing UV to radio coverage. GAMA is funded by the STFC (UK), the ARC (Australia), the AAO, and the participating institutions. This research includes computations using the computational cluster Katana supported by Research Technology Services at UNSW Sydney. 



\section*{Data Availability}

The results of the data presented in the figures are available upon request from the corresponding author. The images are from the Hyper-Suprime Cam Public Data Release 2 \citep[HSC-PDR2;][]{Aihara2019} and can be obtained from \url{https://hsc-release.mtk.nao.ac.jp/doc/index.php/tools-2/} or using the \texttt{unagi} Python module (\url{https://github.com/dr-guangtou/unagi}). The catalogue information was extracted from the Galaxy And Mass Assembly survey (GAMA) Galaxy Group Catalogue \citep[G$^3$Cv10;][]{Robotham2011}. For GAMA data access information we refer the reader to \citet{Driver2022}. 
 





\bibliographystyle{mnras}
\bibliography{All_references_papers}{}







\bsp	
\label{lastpage}
\end{document}